\documentclass[11pt,a4paper]{article}
\usepackage{amssymb}
\usepackage{epsf}
\def\tvi{\vrule height 12pt depth 6pt width 0pt}
\def\tv{\tvi\vrule}
\def\cc#1{\kern .7em\hfill #1 \hfill\kern .7em}

\def\ZZ{\hbox{\it Z\hskip -4.pt Z}}

\begin{document}
\begin{titlepage}

\begin{flushright}
Orsay  LPTHE-96-85 \ \\
Saclay  SPhT-T96/050 \\
Strasbourg LPT-96-21 \\
\end{flushright}

\vskip.5cm
\author{The Author}
\title{The Title }
\date{The Date }

\begin{center}
{\LARGE {\bf Phenomenology of supersymmetric models with a singlet
\footnote{Supported in part by the EC grant Flavourdynamics.}}}
\vskip1.5cm

\medskip

 {\bf U.\ Ellwanger}

LPTHE, Universit\'{e} de Paris-Sud, F-91400 Orsay

\medskip

{\bf M.\ Rausch de Traubenberg}

Laboratoire de Physique Th\'{e}orique, Universit\'{e} Louis Pasteur,

Strasbourg

\medskip

{\bf C.A. Savoy}

Service de Physique Th\'eorique, C. E. de Saclay, F-91191 Gif-sur-Yvette 
\end{center}

\medskip\ \bigskip

\begin{abstract}
\vskip 3pt
The supersymmetric extension of the standard model with an additional
gauge singlet is analysed in detail in the 
light of the recent experimental bounds on supersymmetric particles. The
useful part of the parameter space and the particle spectrum are 
displayed. We find that once the recent bounds on the chargino mass 
are imposed, all other new particles practically satisfy the present
experimental limits. Special attention is given to particles to be searched for
in the future experiments. The singlet fields tend to decouple 
and give rise to an effective MSSM, enlarging the validity of many 
phenomenological analyses based in the minimal field content. However, 
in some ranges of the parameters the singlino is the lightest
neutralino, which modifies the signature for susy particles. Simple
analytical approximations are developed that qualitatively  explain the
numerical results.\- 
\end{abstract}

\end{titlepage}

\section{{Introduction}}

The addition of a $SU(3)\times SU(2)\times U(1)$ gauge singlet
supermultiplet to the MSSM (minimal supersymmetric extension of the
Standard Model) has been motivated by several issues: (i) the solution of
the so-called $\mu $-problem; (ii) the possibility of spontaneous breaking
of the CP symmetry; (iii) the upper limits on the Higgs mass that are
relatively low in the MSSM could be relaxed;
(iv) the neutral particle spectrum is enriched by two scalars and one
fermion, which could mix with other Higgs bosons and neutralinos and
modify their physical properties; (v) possible deviations from the MSSM
could urge    
to enlarge the experimental searches, which have been centred upon the
MSSM.  We refer here to the supersymmetric model with a gauge singlet
in its simplest version, as defined below, as the (M+1)SSM.

However, according to our analysis,
the strong constraints from LEP experiments, when combined
with theoretical and phenomenological requirements, select the 
(M+1)SSM parameters in such a way that the gauge singlet scalars and 
fermions are basically decoupled and the rest of the spectrum 
effectively reproduces a restricted class of the MSSM. Because of this
similarity,  the motivations (ii)-(v) fade out as these issues are excluded 
or diminished after such a detailed investigation. Nevertheless, 
this  also means that phenomenological analyses of the MSSM 
apply to a large extent to the (M+1)SSM as well!
Still, if supersymmetry is found, one can go further and look for some
peculiarites that could distinguish the two kind of models, specially
in the case in which the lightest supersymmetric particle is the singlet
fermion, or singlino. The (M+1)SSM can also be experimentally tested from
the fact that the effective parameters in the equivalent MSSM are very
constrained. 

The present paper is devoted to a thorough discussion of the model in
its version with only dimensionless couplings in its superpotential, 
and universal boundary conditions for the soft supersymmetric breaking
terms at the GUT scale. With these assumptions, the dimensionality of
the parameter  space of the (M+1)SSM equals that of the MSSM. 
Our aim is to understand the pattern of the particle spectrum,
correlations among the particle masses, and features which are
in common with or different from the MSSM. An analysis of the complete
parameters space of the model requires numerical methods, as a scanning over 
the independent parameters at the GUT scale, their subsequent evolution
using the RGE  down to the weak scale, the minimization of the Higgs potential,
the implementation of theoretical and experimental constraints and
finally the computation of the particle masses and some mixing angles.  

Many phenomenological analyses of the  model have appeared
before  \cite{Ellis,ERS1,King,King2,ERS2}. The purpose of the present
work is 
twofold. First, we implement the most recent experimental constraints on
sparticles, Higgs and top quark masses from the LEP and the Tevatron
\cite{PDG,Aleph}. Second, we explain   
the results of our numerical procedure with the help of analytic
approximations  and inequalities among the parameters. Once the vacua
which break colour, electromagnetism or correspond to $\tan \beta
=\infty$ are  
forbidden and the experimental constraints are imposed, these formulae 
allow us to understand the essential features of 
the bounds on and correlations among some particle masses. Quite a lot of 
attention is paid to the property found in \cite{ERS1}, namely, 
that the singlet scalar acquires a v.e.v. which is much larger than the 
Higgs ones. Indeed, this has the important consequence that the singlet 
fermion and scalars quite decouple from the the other multiplets as already 
mentioned.

In section \ref{sec:properties} we review the motivations for the model, 
its essential differences with respect to the MSSM due to the enlargement 
of Higgs sector and the status of the spontaneous CP violation.
The universality assumption for the soft terms and some of its
consequences are explained in section
\ref{sec: universal}. Section \ref{sec:mass} contains the minimum conditions
for the scalar potential as well as the scalar Higgs masses with the
radiative 
corrections included in an approximation that is generally good,
though we use the full expressions \cite{Elrc} in our numerical
calculations.  In section \ref{sec:stability} 
we derive some approximate constraints from the stability of 
the vacuum to understand the pattern of the physical predictions. 
In  section \ref{sec:inos} we discuss some
correlations in the mass spectrum along the same approximations. Then we
summarize our implementation of the 
experimental and theoretical constraints on the model, in section
\ref{sec:cuts}. The main results of the careful scanning of the  parameter 
space are displayed in the figures in section \ref{sec:spectrum}, 
with some emphasis on the sparticles more likely
to be discovered at the colliders, with the exception of the Higgs
sector that has been already presented \cite{ERS2}. The results are
summarized in the last section.
Some useful solutions of the RGE for the soft terms (with generic
boundary conditions, for completeness) are listed in the Appendix.

\vskip 1truecm

\section{{General properties of the (M+1)SSM}}
\label{sec:properties}

The superpotential of the (M+1)SSM (where family mixing is neglected and
irrelevant for the analysis in this paper) is given by 

\begin{eqnarray}
W &=&\lambda SH_{1}H_{2}+\frac{\kappa }{3}S^{3}+
h_{t}TH_{2}Q_3+h_{b}BH_{1}Q_3+h_{\tau }EH_{1}L  \nonumber \\
&&+\mathrm{(similar\ terms\ for\ the\ lighter\ quarks\ and\ leptons)} \label{A}
\end{eqnarray}

\noindent where the chiral superfields are denoted as follows: $S$
(singlet), $H_{1}$ and $H_{2}$ (Higgs doublets), $T$ (antitop), $B$
(antibottom), $E$ (antitau), $Q_3$ (top-bottom doublet), $L$ (tau-neutrino
doublet). Let us briefly review the present status of these motivations for
the (M+1)SSM.

A supersymmetric mass term $\mu H_{1}\cdot H_{2}$ is added to the MSSM
superpotential in order to avoid a potentially dangerous Peccei-Quinn
symmetry and to allow for an acceptable minimum of the Higgs potential. 
The analysis of the $SU(2)\times U(1)$ breaking then requires $\mu $ to be
roughly of the same order of magnitude as the soft terms that embody the
supersymmetry breaking effects in the two Higgs sectors. This interesting
hierarchy problem, the so-called $\mu $-puzzle, has motivated many a
suggestion to link the supersymmetric mass $\mu $ to supersymmetry breaking
\cite{Giuma}.\ In any instance, some kind of new physics is requested and in
this sense the MSSM that looks mostly economical at low energies is not
necessarily so at the unification $(GUT)$ level.\ A natural solution \cite
{NSW1} is provided by the addition of a singlet $S.$ Soft terms that break
supersymmetry can induce a v.e.v. for $S$ and an effective supersymmetric
mass $\mu =\lambda S$ of the right order magnitude. This 
mechanism asks for the $\kappa S^{3}/3$ self-coupling in the superpotential
which also breaks the unwanted Peccei-Quinn symmetry. However, this coupling
would be forbidden and the singlet model inoperative if $S$ has
non-trivial quantum numbers with respect to broken symmetries beyond those
of the Standard Model. This phenomenon is frequently found in superstring
compactifications. We take here the pragmatic viewpoint that all the
dimension three interactions that preserve the low-energy local and global
symmetries are to be included. As we shall discuss later, very low values of
the coupling $\kappa $ are consistent with the phenomenological constraints.
In this case, it could be a relic of non-renormalizable interactions after
the decoupling of, $e.g.,$ extra $U(1)$ gauge symmetries. Actually, 
the ``sliding'' singlet v.e.v. compensates for the smallness of the
couplings $\kappa $ and $\lambda $ and this considerably reduces the
fine-tuning problem of the $\mu $-term. Such a possibility is indeed
supported by our analysis. 

Supersymmetric mass terms are forbidden by the $\ZZ_{3}$ symmetry of the
purely cubic superpotential assumed in (\ref{A}). This discrete
symmetry also prevents $S$ from getting a large v.e.v. through large
radiative corrections to singlet tadpole terms, which invalidate 
other supersymmetric models with gauge singlets \cite{Posus,Elba}.
On the other hand, the $\ZZ_3$ symmetry can give rise to cosmological problems
in the form of domain walls which are produced during the electroweak phase
transition \cite{Abel}. Possible solutions to this problem  have been discussed
in \cite{Abel} and include inflation of the weak scale, embedding of the
discrete symmetry into a gauge symmetry at  the Planck scale or
re-introducing of the $\mu$-term in the superpotential. (In the presence
of the singlet, $\mu$ can be orders of magnitudes smaller than the weak
scale, in which case our subsequent analysis is not affected; however,
the  $\mu$-problem was part of the motivations of the model.) In any
case, since we concentrate  
in this paper on the particle physics aspects of the model and do not
discuss cosmological issues, our results will be independent of the form
of the solutions of the domain wall problem.   

The study of spontaneous CP breaking in the (M+1)SSM also differs from 
the MSSM. Indeed, CP violating vacua can be induced only if
there are more than one phase-dependent interaction in the scalar
potential.\ In the MSSM, the only tree-level term is the soft scalar mass
term $B\mu H_{1}H_{2}$. Analytic terms of higher dimensions generated at 
the one loop level can conspirate with the tree-level one to induce
spontaneous CP violation. However, this has been shown \cite{Poma1} to entail 
a very low upper limit on one of the scalar masses which is inconsistent 
with present experiments at LEP.
The classical scalar potential processes a richer structure in the (M+1)SSM,
with three additional complex scalar couplings (one of dimension four, two
cubic soft terms, see next section). The conditions for CP spontaneous
breaking in this type of models have been shown \cite{Romao} to imply a 
negative eigenvalue in the scalar (mass$)^2$ matrix. 
However, top quark-squark loop corrections could
shift this eigenvalue to positive values. Thus spontaneous CP violation
would occur for appropriate values of the parameters and one or two light
scalars are predicted \cite{Baba}. This issue certainly deserves a more
detailed comparison with phenomenology; it lies outside the purview
of the present investigation. Because the very small region in the (M+1)SSM
parameter space consistent with spontaneous CP violation and those
selected by our CP conserving solutions are apart, we disregard {\it ab
initio} the question of non-trivial phase vacua.

Upper bounds have been derived for the mass of the lightest Higgs scalar in
the MSSM that take into account the large one loop corrections \cite{Haber}
as well as some two-loop effects \cite{ceqw}. This prediction is of
paramount importance in view of the experimental searches of the Higgs
scalar in the LEP 200 and the LHC. The (M+1)SSM has an additional
contribution $\propto \left| \lambda \right| ^{2}$ to the Higgs (mass$)^2.$
It has been shown \cite{Ellis,King,Bine} that requiring the
theory to remain perturbative up to the GUT scale reduces this additional
contribution to $<O$(30 GeV). The upper bounds on the lightest Higgs boson
mass are considerably lowered if one also imposes all phenomenological
constraints needed for the model to be realistic. This has been done in the
case of universal supergravity couplings \cite{ERS1,King}
and also for some special non-universal models \cite{Brax}. Note that a
relatively small increase in the Higgs mass may affect the experimental
signatures. Since the predictions of the phenomenologically constrained 
(M+1)SSM for the Higgs sector has been the subject of a recent publication 
\cite{ERS2}, we shall only give a short account of this important matter
here.

As for the particle spectrum, this model presents several new features as
compared to the MSSM. The supermultiplet $S$ consists of two scalars of
opposite CP and one Majorana fermion. After the $SU(2)\times U(1)$
breaking, these states get mixed with the neutral states in the Higgs
doublet and gauge supermultiplets, leading to a slightly richer
spectrum. In particular, since the visibility of physical states depends
on their 
couplings to the gauge bosons and to matter, some particles could be
relatively light if they are mainly gauge singlets. Thus, some of the limits
on neutral particles obtained at LEP in the MSSM framework do not
immediately apply to the alternative (M+1)SSM. The discussion in this
paper as well as in our previous ones \cite{ERS1,ERS2} is also aimed to
persuade experimentalists to analyze their data on a more general basis. 

In the $\kappa \rightarrow 0,\lambda \rightarrow 0$ limit, the gauge singlet
decouples and the remaining theory looks like the MSSM \cite{Ellis}.
However, this limit leads to a particular version of the MSSM with a
strong correlation among the soft parameters, as we shall discuss later
on. The singlet scalar takes a very large v.e.v. of $O(m_{3/2}/\kappa)$,
but the fermion singlet and both scalar singlets are rather light.

Of course, the large dimension of the scalar field manifold allows for a
rich structure in the overall scalar potential and leads to the existence of
many local extrema \cite{Desa}. One has to carefully check that the phenomenologically viable vacuum is favoured with respect to 
the unwanted ones. Among these there are the usual 
$SU(3)_{c}\times U(1)_{\mathrm{em}}$ breaking ones,
corresponding to vacua with quark and lepton quantum numbers \cite{Frere}, 
\cite{Desa}, \cite{Zwi}. In the (M+1)SSM there is the additional
possibility of a non-vanishing charged Higgs field vacuum.\ However, the
necessary condition for this particular charge violating classical solution
is a relatively large coupling of the Higgses to the singlet \cite{Pool}, so
that this situation is quite naturally avoided by a bound on that coupling.\
Our theoretical and experimental constraints turn out to automatically
selected the right region of the parameter space.

Actually, as we shall discuss here below, an acceptable minimum of the
potential 
requires some inequalities among the soft supersymmetry breaking terms which
would tend to induce $SU(3)_{c}\times U(1)_{\mathrm{em}}$ breaking vacua as
well.\ The radiative corrections due to the gauge interactions then allow
for a compromise while restricting the useful region in the parameter space.
Since the electroweak gauge symmetry breaking is induced by radiative
corrections involving the top Yukawa coupling -- a well-known property in
supersymmetric 
extensions of the Standard Model -- the existence of a physically acceptable
vacuum in the (M+1)SSM is a purely quantum effect.

\vskip 1truecm

\section{{Soft terms and boundary conditions}}\label{sec: universal}

The supersymmetric part of the scalar potential is obtained from the
superpotential through the well-known expression as a sum of $F$ and $D$
terms. The supersymmetry breaking part of the lagrangian contains all
soft terms consistent with the symmetries of the superpotential: gaugino
masses, trilinear analytic scalar couplings and scalar mass terms.\ In the 
(M+1)SSM they are as follows (with the same notation for the chiral
matter supermultiplets and their first component complex scalars) : 
\begin{eqnarray}
\mathcal{L}_{\mathrm{soft}}= 
&&\left( M_{1}\lambda _{1}\lambda _{1}+M_{2}\lambda _{2}\lambda
_{2}+M_{3}\lambda _{3}\lambda _{3}+\right.   \nonumber \\
&&\left. +A_{\lambda }\lambda SH_{1}\cdot H_{2}+
A_{\kappa }\frac{\kappa }{3}S^{3}+A_{t}h_{t}Q_{3}\cdot
H_{2}T^{c}+h.c\right) +\nonumber \\
&&+m_{1}^{2}\left| H_{1}\right| ^{2} +m_{2}^{2}
\left| H_{2}\right| ^{2}  
+m_{S}^{2}\left| S\right| ^{2}+m_{Q_3}^{2}\left| Q_3\right|
^{2}+m_{T}^{2}\left| T^{c}\right| ^{2}+\nonumber \\
&& +... \label{B}
\end{eqnarray}
\noindent where $\lambda _{1},\lambda _{2},\lambda _{3}$ are the gauginos
associated to the $U(1)_{Y},SU(2)$ and $SU(3)$ gauge symmetries. Only the
terms involving the top scalars have been retained in (\ref{B}), with $%
Q_3=(T,B),$ but similar interactions for all other squarks and sleptons 
are to be understood. All the parameters are taken to be real 
(up to irrelevant phases) and flavour
mixing is neglected. Notice that these soft terms contain all the
dimensionful parameters of the model, which are all naturally related to the
supersymmetry breaking scale $M_{\mathrm{susy}}$. In order to study the
particle spectrum and $SU(2)\times U(1)$ symmetry breaking in the (M+1)SSM,
one should take the parameters in (\ref{A}) and (\ref{B}) at the relevant
scales, the Fermi scale $v=174$ {GeV} or $ M_{\mathrm{susy}}.$
Instead, it is natural to fix these parameters at the level of the
underlying supergravity theory. We shall consider here the unification
scale $\Lambda_{\mathrm{GUT}}\sim 10^{16}$ GeV to define them. This is
just a few 
orders of magnitude below the Planck mass and one can presumable assume an
effective renormalizable theory at $\Lambda_{\mathrm{GUT}}.$ Once the
parameters 
are given at $\Lambda_{\mathrm{GUT}},$ their corresponding values at $v$
or $M_{%
\mathrm{susy}}$ are calculated by the integration of the renormalization
group equations (RGE) for the (M+1)SSM \cite{Desa}. We present in the 
appendix the analytic solutions of the (M+1)SSM RGE in the
approximation $h_{t}^{2}\gg  \lambda ^{2},\kappa ^{2},$ which turns out to be
good for most of the set of parameters that will be selected on a
phenomenological basis. However, for the sake of precision, we numerically
integrate the RGE in our analysis, the expressions in the Appendix
being reserved for analytical discussions of the numerical results.
In practice, we renormalize the parameters down to the scale $v$ 
through the supersymmetric RGE, and then apply the Coleman-Weinberg 
corrections to the Higgs potential to take into account the decoupling
of the stops  at $M_{\mathrm{susy}}$. 

The important hypothesis to be made in this paper is the so-called
``universality'' or flavour-independence for the soft-terms in (\ref{B}).
For the supersymmetry breaking scalar interactions it amounts to assume a
common value for all the parameters $A_{i}$ in (\ref{B}) $\mathrm{(at\ }%
\Lambda _{\mathrm{GUT}}\mathrm{)},$

\begin{equation}
A_{\lambda }=A_{\kappa }=A_{t}=A_{b}=A_{\tau }=...=A_0,  \label{C}
\end{equation}

\noindent and the same values (at $\Lambda_{\mathrm{GUT}}$) for the
scalar masses in (\ref{B}),   

\begin{equation}
m_{1}^{2}=m_{2}^{2}=m_{S}^{2}=m_{Q_3}^{2}=m_{T}^{2}=...=m_0^2.  \label{D}
\end{equation}

\noindent This universality follows if the direction of supersymmetry
breaking, characterized by the goldstino components, corresponds to fields
with purely gravitational couplings to the relevant chiral superfields (for
instance, if it is along the dilaton-axion direction in superstring inspired
theories \cite{BFS}). Of course, it follows in general if the goldstino
components have equal couplings to all relevant chiral superfields; an
example is given by the ``large radius limits'' of orbifold
compactifications in string theory if supersymmetry breaking occur along the
modulus direction \cite{Iba}.

As for gaugino masses, the universality hypothesis at
$\Lambda_{\mathrm{GUT}}$ reads   

\begin{equation}
M_{1}\left( \Lambda _{\mathrm{GUT}}\right) =M_{2}\left( \Lambda _{\mathrm{GUT%
}}\right) =M_{3}\left( \Lambda _{\mathrm{GUT}}\right) =M_{0}.  \label{E}
\end{equation}

In grand unified gauge theories many of these relations will be required by
the enlarged gauge group at the tree-level.\ At the one loop level one
expects some threshold corrections to the universality of the soft terms
already at the reference scale $\Lambda _{\mathrm{GUT}}.$ The deviations
from (\ref{C}) - (\ref{E}) in the matter sector can be phenomenologically
important because of the existent experimental bounds on FCNC effects but
they are less relevant to our discussion in this paper.

Deviations from the universality assumption are also possible in the
framework of superstring theory, since the moduli sector can couple
differently to the various chiral fields. Since the assumptions $A_{\kappa
}(\Lambda _{\mathrm{GUT}})=A_{\tau }(\Lambda _{\mathrm{GUT}})$ and $%
m_{1}^{2}(\Lambda _{\mathrm{GUT}})=m_{2}^{2}(\Lambda _{\mathrm{GUT}%
})=m_{S}^{2}(\Lambda _{\mathrm{GUT}})=m_{E}^{2}(\Lambda _{\mathrm{GUT}%
})=m_{L}^{2}(\Lambda _{\mathrm{GUT}})$ play an important role in our
phenomenological analysis, some of our results depend on the universality of
the supersymmetry breaking parameters. However a similar analysis of the 
(M+1)SSM has been published \cite{Brax} for the case of non-universal soft
terms in the framework of an orbifold compactification of the
superstring, and these results are quite analogous to those presented
here. 

The most important consequences of the flavour independence of soft terms
concern the $SU(2)\times U(1)$ symmetry breaking. It could be induced by a
negative value of $m_{2}^{2}$, but for the potential to be bounded from below
and to avoid colour or $e.m.$ symmetry breaking the universal parameter
$m_0^2$ has to be positive,  so that 
$m_{2}^{2}(\Lambda _{\mathrm{GUT}})>0$. For this reason, 
$m_{2}^{2}(v)<0$ has to be enforced by renormalization effects.\ This is
a well-known feature of the MSSM as well, entailing a lower limit on the
top Yukawa coupling. In the (M+1)SSM, the sign of $m_{2}^{2}$ could be made
to flip by the coupling of the singlet fields even for a light top quark 
\cite{Desa}. The experimental value of top quark mass is perfectly
consistent, however, 
with a negative $m_{2}^{2}$ at low energies from the effects of top-stop
quantum loops.

Furthermore the cubic soft terms can also induce spontaneous symmetry
breaking if the value of the parameter $A$ in (\ref{C}) is large enough. 
Hence large values of $A_{\kappa }$ are preferred to give the singlet scalar
a v.e.v..  However, with
the universality assumption (\ref{C}), the squark and sleptons could also
get non-vanishing v.e.v.'s \cite{Frere,Desa,Zwi}. 
Interestingly enough, the gauge interactions reduce the $A$-parameters 
for matter fields while preserving the value of $A_{\kappa },$ as
can be seen from the expressions in the Appendix. Thus, phenomenologically
consistent (M+1)SSM's with the condition (\ref{C}),(\ref{D}),(\ref{E}), 
can be build at
the price of restricting the space of the supersymmetry breaking
parameters.\ This is the most important restriction imposed on the model by
the universality hypothesis, as compared to the MSSM \cite{ERS1}.
Otherwise, the effects of this universality on the $SU(2)\times U(1)$
spontaneous breaking are similar to those in the MSSM.

Another consequence of this assumption is to reduce the number of free
bare parameters in the (M+1)SSM to five (before imposing the physical
constraints) just like in the MSSM: the dimensionful parameters $B$ and
$\mu $ of the 
latter are replaced in the former by the dimensionless parameters
$\lambda $ and $\kappa$. 

At low energies, the soft terms as obtained from the RGE depend on the
parameters $M_{0},A_{0},m_{0}^{2},$ on the choice of the GUT scale, $\Lambda
_{\mathrm{GUT}},$ through $t=$ $\ln \left(\Lambda _{\mathrm{GUT}}/v
\right)/\left(16\pi ^2\right) $ (a typical value being $t\simeq .21),$ and
on the gauge and Yukawa couplings.  
This calculation has to be done numerically
in general. However, from the dimensionality of these running parameters, 
they can be generically expressed at low energy in terms of their 
universal initial values as

\begin{eqnarray}
A_i(t)&=&a_iA_0+b_iM_0 \nonumber \\
m_i^2(t)&=&z_im_0^2+x_iA_0^2+y_iA_0M_0+w_iM_0^2 \label{obvious}
\end{eqnarray}
\noindent where the coefficients depend on the various gauge 
and Yukawa couplings. 
For the simple case where the top Yukawa coupling dominates 
over the others, the dependence on the latter reduces to simple expressions 
in terms of the ratio $\rho =h_{t}^{2}(v)/h_{\mathrm{crit}}^{2}$ where 
$h_{\mathrm{crit}}\simeq 1.1$ is the fixed point value of the top 
Yukawa coupling. The results of this approximation are presented in the 
Appendix and are used for the sake of some qualitative understanding of the 
numerical results.  With the recent measurements
of the top quark mass, $\rho $ is bounded by $\rho \gtrsim 2/3$ (this is
higher than the minimum value required to induce mediative $SU(2)\times U(1)$
breaking). As an example, we list 
the low energy  soft parameters at the fixed point value $\rho =1.$ 
The coefficients are approximated to a reasonable precision.

\begin{eqnarray}
A_{\kappa }&\simeq &A_{0} \ \ \ \  
A_{\lambda } \simeq \frac{1}{2}\left( A_{0}-M_{0}\right) \ \ \ \ 
A_{t} \simeq 2M_{0} \nonumber \\
m_{1}^{2}&\simeq &m_{0}^{2}+\frac{1}{2}M_{0}^{2} \ \ \ \ 
m_{2}^{2}\simeq -\frac{m_{0}^{2}}{2}-3M_{0}^{2} \ \ \ \ 
m_{S}^{2}\simeq m_{0}^{2} \nonumber \\
m_{T}^{2}&\simeq &4M_{0}^{2} \ \ \ \ \ \ \  
m_{Q_3}^{2}\simeq \frac{1}{2}m_{0}^{2}+6M_{0}^{2} \label{crit}
\end{eqnarray}

\vskip 1truecm

\section{{Mass spectrum in the Higgs sector and vacuum structure}}
\label{sec:mass}

Let us first restrict the discussion to the physically relevant vacua with
(real) v.e.v.'s  for the field $\left\langle
H_{2}^{0}\right\rangle =h_{2},\left\langle H_{1}^{0}\right\rangle
=h_{1},\left\langle S\right\rangle =s.$ The scalar potential restricted to
this sector reads $\left( \overline{g}^{2}=g_{1}^{2}+g_{2}^{2}\right) $

\begin{eqnarray}
V &=&\left( \kappa S^{2}+\lambda H_{1}^{0}H_{2}^{0}\right) ^{2}+\lambda
^{2}\left| S\right| ^{2}\left( \left| H_{1}^{0}\right| ^{2}+\left|
H_{2}^{0}\right| ^{2}\right) +  \nonumber \\
&&+\left( A_{\lambda }\lambda SH_{1}^{0}H_{2}^{0}+A\kappa \frac{\kappa }{3}%
S^{3}+h.c.\right) +m_{1}\left| H_{1}^{0}\right| ^{2}+m_{2}^{2}\left|
H_{2}^{0}\right| ^{2}+\nonumber \\
&&+m_{S}^{2}\left| S\right| ^{2}  
+\frac{\overline{g}^{2}}{4}\left( \left| H_{1}^{0}\right| ^{2}-\left|
H_{2}^{0}\right| ^{2}\right) ^{2}+V_{\mathrm{rad}}  \label{F}
\end{eqnarray}

\noindent where $V_{\mathrm{rad}}$ is the quantum correction to the
effective scalar potential beyond the RGE effects included in the running
parameters of (\ref{F}). These radiative corrections are important, in
particular in the evaluation of the lightest Higgs boson mass. The
relevant contribution comes from the top-stop sector 
because of the relatively large value of the top Yukawa coupling. Hence we
only include the top quark and the two stop states, $T_{1}\ \mathrm{and}%
\ T_{2}$ in the usual Coleman-Weinberg expression,

\begin{equation}
V_{\mathrm{rad}}=\frac{1}{64\pi ^{2}}\;STr\;m^{4} \ln \frac{m^{2}}{%
Q^{2}}  \label{G}
\end{equation}

The appropriate mass terms in (\ref{G}) are then the top quark mass
$m_t^2$  and the stop masses 

\begin{eqnarray}
m_{T_{1,2}}^{2} &=&m_{t}^{2}+\frac{1}{2}\left(
m_{Q_{3}}^{2}+m_{T}^{2}\right) \pm W, \mathrm{where}  \nonumber \\
W^{2} &=&\frac{1}{4}\left( m_{Q_{3}}^{2}+m_{T}^{2}\right) +h_{t}^{2}\left|
A_{t}h_{2}+\lambda sh_{1}\right| ^{2},  \nonumber \\
m_{t} &=&h_{t}h_{2}.  \label{H}
\end{eqnarray}

All the parameters in (\ref{F}) and (\ref{G}) are to be taken at the scale $%
Q^{2}\sim O\left( h_{1}^{2}+h_{2}^{2}\right) .$

\noindent These radiative corrections are mostly due to the splitting
between the quarks and the squarks.
However, even if one starts from universal parameters
at $\Lambda_{\mathrm{GUT}},$ the RGE running introduce some asymmetry in the
stop sector as can be seen from the formulae in the Appendix. The resulting
splitting in $m_{T_{1}}^{2}-m_{T_{2}}^{2}$ produces a sizeable effect in $V_{%
\mathrm{rad}}$ and in the scalar mass spectrum.\ The full one loop
expressions for the mass matrices and the potential minimization are
available in the literature. All the numerical calculations in the present
paper are performed by using the formulae in \cite{Elrc}. However, for the
sake of the analytic approximations discussed below we use a simple
approximation by keeping only the lowest relevant order in the development
in terms of the stop splitting,
$\left( m_{T_1}^2-m_{T_2}^2\right) /\left( m_{T_1}^2+m_{T_2}^2\right) .$ 
The minimization of the scalar potential amounts to the following conditions

\begin{equation}
\begin{tabular}{lll}
$h_{1}\left[ m_{1}^{2}+\hat{\mu}^{2}\left( 1+\frac{h_{2}^{2}}{s^{2}}\right) +%
\frac{1}{2}\overline{g}^{2}\left( h_{1}^{2}-h_{2}^{2}\right) +\right. $ &  & 
\\ 
&  &  \\ 
\multicolumn{1}{r}{$\left. +\hat{B}\hat{\mu}\frac{h_{2}}{h_{1}}+\beta _{t}%
\hat{\mu}^{2}L\right] $} & $=$ & $0$ \\ 
&  &  \\ 
$h_{2}\left[ m_{2}^{2}+\hat{\mu}^{2}\left( 1+\frac{h_{1}^{2}}{s^{2}}\right) -%
\frac{1}{2}\overline{g}^{2}\left( h_{1}^{2}-h_{2}^{2}\right) \right] +$ &  & 
\\ 
&  &  \\ 
$+\hat{B}\hat{\mu}\frac{h_{1}}{h_{2}}+\beta _{t}\left[ \left(
m_{T}^{2}+m_{Q_{3}}^{2}+A_{t}^{2}\right) \left( L+\varepsilon \right)
\right. $ &  &  \\ 
&  &  \\ 
\multicolumn{1}{r}{$\left. -\left( m_{T}^{2}+m_{Q_{3}}^{2}\right)
+2m_{t}^{2}L\right] $} & $=$ & $0$ \\ 
&  &  \\ 
$s\left[ m_{S}^{2}+\hat{\mu}\frac{h_{1}^{2}+h_{2}^{2}}{s^{2}}+2\nu
^{2}+A_{\kappa }\nu +\left( \hat{B}+\nu \right) \hat{\mu}\frac{h_{1}h_{2}}{%
s^{2}}+\right. $ &  &  \\ 
&  &  \\ 
\multicolumn{1}{r}{$\left. +\beta _{t}\hat{\mu}^{2}\left( \frac{h_{2}}{s}%
\right) ^{2}\left( L+\varepsilon \right) \right] $} & $=$ & $0$%
\end{tabular}
\label{I}
\end{equation}

\noindent where

\begin{equation}
\beta _{t} =\frac{3h_{t}^{2}}{16\pi ^{2}}\ ,\ \ \ \ \   
\varepsilon =\ln \frac{%
m_{t}^{2}}{Q^{2}}\ ,\ \ \ \ \    
L =\ln \frac{m_{T_{1}}m_{T_{2}}}{m_{t}^{2}}  \label{J}
\end{equation}

\noindent and the variables

\begin{eqnarray}
\hat{\mu} &=&\lambda s  \nonumber \\
\nu &=&\kappa s  \nonumber \\
\hat{B} &=&A_{\lambda }+\nu +\beta _{t}A_{t}\left( L+\varepsilon \right)
\label{K}
\end{eqnarray}

\noindent have been introduced. An one loop term has been
conveniently included in the definition of the effective parameter $\hat
B$ (of the MSSM). 

The $3\times 3$ mass matrix for the CP even scalars (in the same
approximation) has the following matrix elements

\begin{eqnarray}
m_{11}^{2} &=&\overline{g}^{2}h_{1}^{2}-\hat{B}\hat{\mu}\left( \frac{h_{2}}{%
h_{1}}\right) -\beta _{t}\hat{\mu}^{2}\frac{Z^{2}}{3}  \nonumber \\
m_{22}^{2} &=&\overline{g}^{2}h_{2}^{2}-\hat{B}\hat{\mu}\left( \frac{h_{1}}{%
h_{2}}\right) +4\beta _{t}m_{t}^{2}L+\beta _{t}A_{t}\left( 2m_{t}Z%
-A_{t}\frac{Z^{2}}{3}\right)  \nonumber \\
m_{SS}^{2} &=&\left( A_{\kappa }+4\nu \right) \nu -\left( \hat{B}-\nu
\right) \hat{\mu}\frac{h_{1}h_{2}}{s_{2}}-\frac{\hat{\mu}^{2}}{3}\left( 
\frac{h_{1}^{2}}{s_{2}}\right) \beta _{t}Z^{2}  \nonumber \\
m_{1S}^{2} &=&2\hat{\mu}^{2}\left( \frac{h_{1}}{s}\right) +\left( \hat{B}%
+\nu \right) \hat{\mu}\left( \frac{h_{2}}{s}\right) +\beta _{t}\hat{\mu}%
^{2}\left( \frac{h_{1}}{s}\right) \left( L+\varepsilon -\frac{1}{2}\right) 
\nonumber \\
m_{2S}^{2} &=&2\hat{\mu}^{2}\left( \frac{h_{2}}{s}\right) +\left( \hat{B}%
+\nu \right) \hat{\mu}\left( \frac{h_{1}}{s}\right) +\hat{\mu}\left( \frac{%
h_{1}}{s}\right) \beta _{t}\left( m_{t}Z-A_{t}\frac{Z^{2}%
}{3}\right)  \nonumber \\
m_{12}^{2} &=&-\overline{g}^{2}h_{1}h_{2}+2\hat{\mu}^{2}\left( \frac{%
h_{1}h_{2}}{s^{2}}\right) +\hat{B}\hat{\mu}+\hat{\mu}\beta _{t}\left( m_{t}%
Z-A_{t}\frac{Z^{2}}{3}\right)  \nonumber \\
Z &=&\frac{m_{t}\left( A_{t}+\hat{\mu}\cot \beta \right) }{%
m_{t}^{2}+\frac{1}{2}\left( m_{Q_{3}}^{2}+m_{T}^{2}\right) }  \label{L}
\end{eqnarray}
The parameter L characterizes the top-stop mass splitting
and can be relatively large, while $\varepsilon $ depends on the low
energy scale chosen to define the running parameters --- we take
$Q^{2}\sim \left( h_{1}^{2}+h_{2}^{2}\right) $ in this paper so that
$\left| \varepsilon \right| \sim \left| \ln h_{t}^{2}\right| <1.$

The CP odd scalar (mass)$^2$ matrix with the would-be Goldstone boson
projected out is as follows,

\begin{equation}
\left( 
\begin{array}{ll}
-\hat{B}\hat{\mu}\frac{h_{1}^{2}+h_{2}^{2}}{h_{1}h_{2}} & -\left( \hat{B}%
-3\nu \right)  \hat{\mu}\frac{\sqrt{h_{1}^{2}+h_{2}^{2}}}{s} \\ 
&  \\ 
-\left( \hat{B}-3\nu \right) \hat{\mu}\frac{\sqrt{h_{1}^{2}+h_{2}^{2}}}{s} & 
-3A_{\kappa }\nu -\left( \hat{B}+3\nu \right) \hat{\mu}\frac{h_{1}h_{2}}{%
s_{2}}
\end{array}
\right)  \label{M}
\end{equation}
The minimization of the potential fixes  
$\hat{B}\hat{\mu}$ and $ A_{\kappa }\nu $ to be negative, so that 
the diagonal entries in (\ref{M}) are positive. The (mass)$^2$ of the 
charged scalar reads 

\begin{equation}
m_{H^{+}}^{2}=M_{W}^{2}\left( 1-\frac{\lambda ^{2}}{g_{2}^{2}}\right)
+\left| \hat{B}\hat{\mu}\frac{h_{1}^{2}+h_{2}^{2}}{h_{1}h_{2}}\right|
\label{N}
\end{equation}
The radiative corrections are incorporated into the CP odd neutral and
charged scalar masses through the shift of the $\hat{B}$ parameter as
defined in (\ref{K}) \cite{Elrc}.

Although all the numerical results in this paper are obtained by numerical
minimization of the complete one loop potential (\ref{F}), (\ref{G}) and the
numerical determination of the mass eigenvalues, it is interesting to get
some insight into the results through simple analytic approximations.
Indeed, many features of the particle spectrum of the (M+1)SSM are related
to an important property, first noticed in \cite{ERS1}, namely:
the only physically acceptable solutions of the vacuum equations (\ref{I})
are large singlet solutions: $ s \gtrsim 1\mathrm{TeV}\gg 
h_{2},h_{1}.$
Indeed, solutions with $ s \lesssim h_{2}$ lead to
one or more light states in the spectrum that are excluded by experiments,
specially by LEP experiments. Examples can be found in \cite{Ellis}, where
the parameters have been chosen to yield light spectrum. The general
character of this feature of the (M+1)SSM has been carefully checked in
our numerical analysis.
Having settled this fact one can go on to establish several approximate
consequences.

(A) The $S$ v.e.v is approximately obtained from the third equation in
(\ref{I}) with $h_2=h_1=0$, 

\begin{equation}
s \simeq \frac{-A_{\kappa }-\sqrt{A_{\kappa }^{2}-8m_{S}^{2}}}{4\kappa }
 \label{O}
\end{equation}
\noindent The potential at this value of $s$ has also to be negative
in the allowed region of the parameter space, as discussed in more detail
in section \ref{sec:stability}. This additional condition requires 
\begin{equation}
9m_{S}^{2}<A_{\kappa }^{2}
\label{neccon}
\end{equation}
\noindent This approximation turns out to be very accurate. 
The parameters $A_{\kappa} $ and $m_{S}^{2}$ are only slightly renormalized 
if $\kappa ,\lambda \ll 1. $ Actually, these relatively low values of 
the couplings are also required by our numerical study of the model, 
so that the condition (\ref{neccon})
 approximately applies to the bare universal parameters $A_{0},m_{0}^{2},$ yielding the strong constraint in the parameter space:
\begin{equation}
A_{0}^{2} \gtrsim 9m_{0}^{2}
\label{neccon1}
\end{equation}

The $S$ v.e.v. in (\ref{O}) defines the effective parameters $\hat{B}$ and 
$\hat{\mu}$ analogous to those often denoted $B$ and $\mu $ in the MSSM.
Indeed, up to corrections of $O\left( h_{2}^{2}/s^{2}\right) ,$ the first
two equations in (\ref{I}) are the same as in the MSSM with the
introduction of these parameters.
Notice that $\hat{\mu}\sim \lambda A_{\kappa }/2\kappa$ should be of 
$O\left( m_{3/2}\right) .$ Hence the well--known ``fine-tuning'' on the  
MSSM free parameter, $\mu \sim O\left( m_{3/2}\right) ,$ is replaced here
by the requirement of no strong hierarchy between the $\lambda $ and $\kappa 
$ Yukawa couplings, as the $SU(2)\times U(1)$ invariant scale of the singlet
field is dynamically fixed by supersymmetry breaking, accordingly to 
(\ref{O}).

With $\tan \beta =h_2/h_1$ one recovers, in this approximation, the MSSM
relation
\begin{equation}
\tan ^{2}\beta =\frac{m_{1}^{2}+\beta_{t}\hat{\mu}^{2}L+\hat{\mu}^{2}+
M_{Z}^{2}/2}{m_{2}^{2}+\left(
m_{T_{1}}^{2}+m_{T_{2}}^{2}+A_{t}^{2}\right) \beta_{t}L  +\hat{\mu}^{2}+ M_{Z}^{2}/2} 
\label{P} 
\end{equation}

Only the dominant terms of the radiative corrections are retained. Because
the denominator in (\ref{P}) has terms of either sign $\left(
m_{2}^{2}<0\right) $ it is relatively sensitive to radiative corrections
which tend to decrease $\tan \beta $ with respect to the tree-level
approximation. 

(B) In the neutral scalar mass matrices (\ref{L},\ref{M}) as well 
as in the neutralino mass matrices, the mixing of the singlet
fields to the others is always proportional to $h_1/s$ and $h_2/s$,
hence small. This is the reason why, in most of the allowed points of the
parameter space, the singlet sector of the theory, after dynamically
producing the effective parameters (\ref{K}), is almost decoupled from
the rest of the theory that resembles to the MSSM.

(C) The MSSM is formally obtained in the limit 
$s\rightarrow \infty $ with $\nu $
and $\hat{\mu}$ fixed $\left( \kappa ,\lambda \ll \overline{g}^{2}\right) .$
In this limit the singlet sector decouples from the doublet Higgs sector,
and the singlet masses are:
\vskip .5 truecm

\begin{eqnarray}
\frac{1}4\sqrt{A_0^2-8m_0^2}\left(\left| A_0\right| 
+\sqrt{A_0^2-8m_0^2}\right)  
\mathrm{(}CP\mathrm{\ even\  scalar)} & \nonumber \\ 
&  \nonumber  \\ 
\frac {3}{4}\left| A_0\right| \left(\left| A_0\right| 
+\sqrt{A_0^2-8m_0^2}\right)  
\mathrm{(}CP\mathrm{\ odd\  scalar)} & \\   & \nonumber \\ 
\left(\left| A_0\right| +\sqrt{A_0^2-8m_0^2}\right) /2  \mathrm{\ ( 
 singlet\ fermion,\ singlino).}& \nonumber 
\end{eqnarray}

\vskip .5 truecm
\noindent The singlet particles become light in the small $A_{0}$ limit. We
argue in section \ref{sec:stability} that this limit is 
always associated to the singlet decoupling limit

\noindent $\left( \kappa \rightarrow 0,s^{2}\gg  h_{1}^{2}+h_{2}^{2}\right) .$
Some of the physical consequences of a light singlino are discussed in section 
\ref{sec:inos}.

The neutral and charged Higgs\footnote{%
As can be seen from (\ref{N}) there is a negative contribution 
$\lambda ^{2}M_{Z}^{2}/\overline{g}^{2},$ to $m_{H^{+}}^{2}.$ This is
the origin of the condition $\lambda ^{2}<\overline{g}^{2}$ which is
sufficient to avoid $e.m.$ charge breaking through $\left\langle
H^{+}\right\rangle \neq 0,$ which is satisfied anyway by the models that are
consistent with the other phenomenological constraints.} sector, as well as
other neutralinos and charginos follow the MSSM pattern. In this sense the 
(M+1)SSM reduces to the MSSM in this limit but with an important
restriction in the parameter space needed to dynamically implement the
physically acceptable vacuum.

\vskip 1truecm
 
\section{{Constraints  on  the   parameter  space from 
vacuum stability}}
\label{sec:stability}

The spontaneous breaking of $SU(2)\times U(1)$ symmetry in the (M+1)SSM
restricts even more the $(M_{0},m_{0}^{2},A_{0})$ parameter subspace than in
the MSSM under the same assumption of universality. The minimum discussed
in section 3 has to be cosmologically flavourless and stable with respect to
other possible minima. The first constraint comes from the fact that the $%
S$ v.e.v. is induced through the ``$A$-mechanism'' \cite{BFS,NSW1}
with $A_{0}^{2}>9m_{0}^{2},$ as given in (\ref{neccon1}). It is well known that
large $A_{0}$ values may induce charged and coloured vacua.

In order to avoid slepton v.e.v.'s  one has to impose

\begin{equation}
A_{\mathrm{e}}^{2}<3\left( m_{E}^{2}+m_{L}^{2}+m_{1}^{2}\right)  \label{R}
\end{equation}

\noindent at a scale $\Lambda \lesssim 0\left( A_{\mathrm{e}}/h_{\mathrm{e}%
}\right) ,$ where $\left( A_{\mathrm{e}}/h_{\mathrm{e}}\right) $
characterizes the scale of the would-be slepton $v.e.v.^{\prime }s.$ From
the RGE solutions in the appendix this implies 
\newline $\left( t_{\mathrm{e}}=\ln \left(
h_{e}\Lambda _{\mathrm{GUT}}/A_{\mathrm{e}}\right) /16\pi ^{2}\right), $
\begin{eqnarray}
&&\left( A_{0}+2M_{0}t_{\mathrm{e}}\left( 3g_{2}^{2}+3g_{1}^{2}\right) \right)
\nonumber \\
&<&9\left[ m_{0}^{2}+2g_{2}^{2}t_{\mathrm{e}}\left( 1-g_{2}^{2}t_{\mathrm{e}%
}\right) M_{0}^{2}+2g_{1}^{2}t_{\mathrm{e}}\left( 1-11g_{1}^{2}t_{\mathrm{e}%
}\right) M_{0}^{2}\right].  \label{S}
\end{eqnarray}

The most dangerous v.e.v.'s comes from the electron sector with the smallest
Yukawa and the largest scale $A_{\mathrm{e}}/h_{\mathrm{e}}\sim 0\left(
10^{8}GeV\right) $, and (\ref{R}) at this scale finally becomes 

\begin{equation}
\left( A_{0}+.4M_{0}\right) ^{2}<9m_{0}^{2}+2.4M_{0}^{2}.  \label{T}
\end{equation}

\noindent This can be combined with the condition $A_{0}^{2}>9m_{0}^{2}$ 
to obtain restrictions on the ratio $A_{0}/M_{0}.$ 
For this purpose we introduce a parameter to characterize the fine-tuning 
of $m_{0}^{2}$ with respect to $A_{0},$

\begin{equation}
\alpha =\frac{A_{0}^{2}-9m_{0}^{2}}{A_{0}^{2}}  \label{U}
\end{equation}

\noindent and obtain from (\ref{T}) the upper bound on $\alpha$ 

\begin{equation}
\left| \frac{M_{0}}{A_{0}}\right| >.18\left( \sqrt{14\alpha +1} + 
\mathrm{sign}(A_0/M_0) \right)  \label{V}
\end{equation}

For $A_{0}/M_{0}>0,$ (\ref{T}) yields the limit

\begin{equation}
M_{0}^{2}>\frac{A_{0}^{2}}{8}>m_{0}^{2}  \label{W}
\end{equation}

\noindent For $A_{0}/M_{0}<0,\left| A_{0}\right| $ can become much larger
than $M_{0}$ at the price of a fine-tuning of $m_{0}^{2}/A_{0}^{2}$ 
corresponding to $\alpha $ of $O\left( M_{0}/A\,_{0}\right).$
Though these conditions are only valid under the universality assumption for
soft terms, analogous ones can be derived along the same lines for other
models (see, $e.g.,$ \cite{Brax}). (Strictly speaking, (\ref{T}) can only be
analytically derived \cite{Desa} if the scalars are relatively degenerate.
But recent numerical studies suggest that it is approximatelly valid 
under more general conditions \cite{bordner})

The next step is to avoid minima with $\left\langle H_{1}\right\rangle =0,%
\mathrm{i.e.,}$ with $\tan \beta \rightarrow \infty .$ In this analysis the
radiative corrections discussed in section 3 can be relevant. Of course, in
our numerical analysis they are included, but in order to understand the
origin of some of the constraints on the (M+1)SSM parameter space, 
they are neglected in some of the expressions here below. In the effective
two-step minimization of the scalar potential, one finds for the
phenomenologically acceptable solution discussed in the previous section 
the approximate value at the minimum

\begin{eqnarray}
V_{\min } &=& \frac{-A_{\kappa }^{4}}{48\kappa ^{2}}C(\alpha )
 -\frac{M_{Z}^{4}}{4\overline{g}^{2}}\cos ^{2}2\beta  \nonumber \\ 
C(\alpha ) &=& \left( \frac{3+\sqrt{1+8\alpha }}{6} \right) ^{3}
\left( \frac{\sqrt{1+8\alpha }-1}{2}\right)  \ \label{X}
\end{eqnarray}

\noindent where the first term comes from the pure singlet sector and the
last one from the effective MSSM potential defined with the parameters
given by (\ref{K}).
This physical minimum has to be compared with another one, with $%
\left\langle S\right\rangle =\left\langle H_{1}\right\rangle =0$ and $%
\left\langle H_{2}^{2}\right\rangle =-m_{2}^{2}/2\overline{g}^{2}.$ 
The potential at this minimum is 

\begin{equation}
V_{\min }^{\prime }=-\frac{(m_{2}^{2})^2}{\overline{g}^{2}}  \label{Y}  
\end{equation}

In principle one should look for the cosmological formation and stability of
the minimum (\ref{X}) with respect to (\ref{Y}). For simplicity we replace
the rigorous constraints by the simpler condition that $V_{\min }<V_{\min
}^{\prime },$ {\it i.e.,} that the absolute minimum is the physical one.\
This stability condition differs from the corresponding one within the 
MSSM in two respects: (i) the presence of the pure singlet negative term in
(\ref{X}) favours $V_{\min };$ (ii) in the MSSM, $(-m_{2}^{2})^{2}$ is
replaced by   
$(-m_{2}^{2}+\hat{\mu}^{2})^{2}$ in (\ref{Y}), so that $V_{\min
}^{\prime }$ is more dangerous in the (M+1)SSM. The comparison of
(\ref{X}) and (\ref{Y}) leads to the condition (at the  scale $v$),

\begin{equation}
\frac{\overline{g}^{2}A_{\kappa }^{4}}{48\kappa ^{2}}
C(\alpha )+\frac{M_Z^{4}}{4}\cos ^{2}2\beta > (m_{2}^{2})^{2}.
\label{Z}  
\end{equation}

On the other hand, from the RGE solution in the appendix one gets

\begin{equation}
-m_{2}^{2}>\left( \frac{3\rho }{2}-1\right) m_{0}^{2}+\left( 4\rho -\frac{1}{%
2}\right) M_{0}^2  \label{A1}
\end{equation}

\noindent and the measured values of the top mass does not allow for
small $\rho = h_{t}^{2}/h_{\mathrm{crit}}^{2}$ so that $\rho \gtrsim
2/3$.  In addition, in the next 
section we derive from the experimental limits on the chargino masses
a bound on $M_{Z}/M_{0},$ 

\begin{equation}
\frac{2M_{0}}{\sqrt{5}} >M_Z=\overline{g}\sqrt{h_{1}^{2}+h_{2}^{2}}\ .
\label{minM0} 
\end{equation}

\noindent (At this point, one is introducing experimental information on
the spectrum. As already stressed, it is the interplay between the 
LEP bounds on sparticle masses and the vacuum conditions that constrains
the parameter space.) Now, from (\ref{minM0}) and (\ref{A1}), 
(\ref{Z}) can be well approximated by

\begin{equation}
\frac{\overline{g}}{4\sqrt{3} \kappa } >\left( \frac{M_{0}^{2}}{A_{0}^{2}}\right) 
\frac{\left( 4\rho -\frac{1}{2}\right) }{\sqrt{C(\alpha )}}  \label{A2}
\end{equation}
\noindent Notice that the necessary condition  $C(\alpha ) >0 $, which
follows from (\ref{minM0}) and (\ref{Z}), implies $\alpha >0,$ namely, the condition (\ref{neccon1}). 

For $A_{0}/M_{0}>0$ one obtains from (\ref{V}) and (\ref{A2}) the bound

\begin{equation}
\frac{\kappa }{\overline{g}}<\frac{1}{5\left( 4\rho -\frac{1}{2}\right)}  
 \ , \label{A3}
\end{equation}

\noindent so that $\kappa $ is always relatively small, $\kappa <\overline{g}%
/10=0.05.$ This is related to the large values taken by the $S$ v.e.v..
Indeed by replacing (\ref{O}) and (\ref{minM0}) in (\ref{A2}),  we obtain

\begin{equation}
\frac{\left| s\right| }{\sqrt{h_{1}^{2}+h_{2}^{2}}}>\sqrt{3} \left( 8\rho -1\right)
>7 \ . \label{A4}
\end{equation}

Of course, the limits (\ref{A3}) and (\ref{A4}) are relatively loose bounds
and most of the solutions of our numerical studies correspond to smaller
(larger) values of $\kappa $ ($s,$ respectively). Also to be noticed is the
behaviour of $s$ for fine-tuning of the parameter $\alpha $ in
(\ref{U}): $s/\sqrt{h_{1}^{2}+h_{2}^{2}}\rightarrow \left( 4\rho
-1/2\right) /\sqrt{\alpha }$ as $\alpha \rightarrow 0,$ while $\kappa $
decreases as $\sqrt{\alpha }.$ 

For $A_{0}/M_{0}<0,$ the situation is more involved since one can allow for
large values of $A_{0}$ with some fine-tuning of the $m_{0}^{2}/A_{0}^{2}$
ratio. The relations corresponding to (\ref{A3}) and (\ref{A4}) are
respectively

\begin{eqnarray}
\frac{\kappa }{\overline{g}} &<&\frac{1}{2 \left( 4\rho -\frac{1}{2} \right) }\frac{1}{%
\sqrt{\alpha }}\left( 1+\frac{1}{6\alpha }\right)  \nonumber \\
\frac{\left| s\right| }{\sqrt{h_{1}^{2}+h_{2}^{2}}} &>&\left( 2\rho -\frac{1}{4} \right) \frac{7\sqrt{\alpha }}{(1+\alpha )}  \label{A5}
\end{eqnarray}

\noindent For moderate values of $\alpha ,$ (\ref{A5}) corroborate our
conclusions from (\ref{A3}) and (\ref{A4}). In the fine-tuning limit, $%
\alpha \rightarrow 0$,  the singular behaviour of $\kappa $ 
in the parameter $\alpha $ is apparent.
However, in this case our approximations are not powerful 
enough to explain the lower bound \cite{ERS1} 
$s\gtrsim 6\sqrt{h_{1}^{2}+h_{2}^{2}} $ that is found in our numerical
analysis.
 
Another interesting constraint follows from (\ref{P}) which implies $
-(m_{2}^{2}+\left( m_{T_{1}}^{2}+m_{T_{2}}^{2}+A_{t}^{2}\right)
\beta_{t}L) 
<\hat{\mu}^{2}+M_{Z}^{2}/2.$ Neglecting the top-stop
radiative corrections and using (\ref{A1}) and (\ref{minM0}) yields, 

\begin{equation}
\hat{\mu}^{2}=\lambda ^{2}s^{2} > \left( 4\rho -1\right) M_{0}^{2}  \label{A6}
\end{equation}

\noindent From (\ref{O}) and (\ref{V}), one derives the limits:
\begin{equation}
\begin{array}{lll}
\left( \lambda ^{2}/\kappa ^{2}\right)  & > &
\left( 4\rho -1 \right) \ \ \ \ \ \ \ \ \ \ \  \ 
\left( A_{0}/M_{0}>0\right) \nonumber \\
{ } & > &
\left( 4\rho -1 \right)\left(3\alpha /(1+2\alpha )\right) \ \ \left(
A_{0}/M_{0}<0\right) \end{array}
 \label{A7} 
\end{equation}

\noindent For moderate values of $\alpha $ this implies $\kappa
^{2}<\lambda ^{2} .$ At the price of some fine-tuning and $%
A_{0}/M_{0}<0$ one can obtain $\kappa $ slightly larger than $\lambda .$ 

Concerning the important parameter $\tan \beta ,$ 
some simple qualitative predictions can be obtained from the following
expression,   

\begin{equation}
\frac{-2\hat{B}\hat{\mu}}{m_{1}^{2}+m_{2}^{2}+\left( m_{T_{1}}^{2}+m_{T_{2}}^{2}+A_{t}^{2}+\hat{\mu}^{2}\right) \beta_{t}L
+2\mu ^{2}}=\sin 2\beta 
\label{Q'}
\end{equation}
\noindent which is equivalent to (\ref{P}) by the minimum conditions (\ref{I}).
Notice that $\mathrm{sign} (\tan \beta)=-\mathrm{sign} (\hat{B} \hat{\mu} )$.
From the renormalized expression (\ref{Z6}) in the appendix, one sees
that for $A_0M_0<0$ the sign of  
$\hat{B}$ is preserved and its magnitude increased by the 
gauge renormalisation proportional to $M_0$ which increases the numerator 
in (\ref{Q'}). Therefore, in this case, $\tan \beta >0$ and it can take 
relatively small values for $\left| A_0\right|  > \left| M_0\right| .$
Instead, for $A_0M_0>0,$ where (\ref{W}) also applies, the $M_0$ term
in (\ref{Z6}) tends to exceed the $A_0$ term so that in most cases 
 $\tan \beta < 0$ and its magnitude is relatively large. (Exceptions
are the special cases with small $\rho $ and 
$\left| A_0 \right| \gtrsim 2\left| M_0 \right| $ with a positive large
$\tan \beta$.)  

Therefore, with these simple analytic approximations for the vacuum
conditions, together with the approximated condition from the experimental
limits on the gauginos (to be discussed later on), one easily understands
the qualitative pattern of the allowed parameter space as obtained from
the detailed numerical investigation. For instance, (\ref{A4}) and (\ref{A5})
are related to the decoupling of the singlet sector, whereas sizeable
mixing - which requires $s/\sqrt{h_{1}^{2}+h_{2}^{2}} \sim O(1)$ - is 
possible only with the fine-tuning (and signs) discussed below (\ref{A5}).
\vskip 1truecm
  
\section{{Qualitative aspects of the mass spectrum}}\label{sec:inos}

Let us first consider the chargino masses given by the expression

\begin{equation}
\begin{array}{lll}
m_{\chi^{+}}^{2} & = & \frac{1}{2}\left( M_{2}^{2}+\hat
{\mu}^{2}+M_{W}^{2}\right) \nonumber \\ 
&  &  \nonumber \\ 
& \pm & \frac{1}{2}\sqrt{\left( M_{2}^{2}+\hat {\mu}^{2}+M_{W}^{2}\right)^{2} -\left( 
2\hat {\mu} M_{2}-M_{W}^{2}\sin 2\beta
\right) ^{2}}
\end{array}
\label{E1}
\end{equation}

\noindent with $M_{2}^{2}/M_{0}^2 \simeq 2/3.$ For the LEP experimental
limit we take 
$m_{\chi^{+}}>65$ {GeV}. Of course, the constraints from this limit 
on the parameters are the same as in the MSSM (with $\tan \beta$
in the more restricted range determined by the (M+1)SSM as discussed in 
section \ref{sec:spectrum}). On the other hand, this constraint 
 plays an important role in the discussion of the previous section as 
well as in this section. 

Because the scale of the soft supersymmetry breaking parameters is set
by the weak scale, we conveniently express the dimensionful parameters
in terms of $M_Z$ in this section. 
One can use the experimental bound, together with (\ref{P}) 
and the (\ref{A1}), to deduce the relations

\begin{equation}
\hat{\mu}^{2}>\left( 6\rho -\frac{3}{4}\right) M_{2}^{2}-
\frac{M_{Z}^{2}}{2}>M_{2}^{2}>m_{\chi^{+}}^{2}> M_{Z}^{2}/2. 
 \label{E2}
\end{equation}
Let us now define a parameter $a$ by,
\begin{equation}
\hat{\mu}^{2}+\frac{M_{Z}^{2}}{2}=\left( a-1\right) M_{2}^{2}  \label{E3}
\end{equation}
\noindent which, for $\tan \beta \gtrsim 2$ 
(which turns out to be always the case in
our numerical analysis), satisfies the approximate inequalities 
$6\rho +1/4\lesssim a\lesssim 6\rho + 2.$ 
Then, from (\ref{E1}) and (\ref{E2}) 
one gets the following expression for the lightest chargino mass,

\begin{equation}
m_{\chi^{+}}^{2}\simeq \frac{\left( a-1\right) }{a}M_{2}^{2}-\left( 1+%
\frac{2\sqrt{a}}{\tan \beta }\right) \frac{3M_{Z}^{2}}{4a} \ , \label{E4}
\end{equation}

\noindent so that the approximate bounds,  

\begin{equation}
\frac{1}{14}M_{3}^{2}-\frac{1}{2}M_{Z}^{2}\simeq 
\frac{3}{4}M_{2}^{2}-\frac{1}{3}M_{Z}^{2} \lesssim m_{\chi^{+}
}^{2}\lesssim \frac{7}{8}M_{2}^{2}-\frac{1}{6}M_{Z}^{2} \simeq
\frac{2}{25}M_{3}^{2}-\frac{1}{6}M_{Z}^{2}  \label{E5}  
\end{equation}
 
\noindent are obtained by varying the parameters $\rho$ and $\tan \beta ,$ 
and by using the relation between the gluino and wino masses 
$M_{3}^{2}/M_{2}^{2} \simeq 10.7.$  This defines a relatively 
narrow band for $m_{\chi^+}$ as a function of the gluino mass $M_3$,
that is reasonably reproduced in our numerical analysis.  
It just expresses the fact that the lightest chargino 
is mostly a wino, with a component of $O(1/a)$ of the higgsino. 

Now, from (\ref{E5}) and the experimental 
LEP bound, $m_{\chi^{+}}^{2}>M_{Z}^{2}/2,$ one gets a limit on the
gluino mass, $M_3 \gtrsim 3M_{Z}= 273$ {GeV} (from the previous
LEP bound, $m_{\chi^{+}}>M_{Z}/2,$ the corresponding result
was $M_{3}\gtrsim 180$ {GeV}, but needed a more detailed approximation
than (\ref{E5})). This corresponds to $M_{0} \gtrsim 100$ GeV. 
Notice that these are only approximations aimed to explain rather than
predict the numerical results that take all the parameters into account. 
As already stressed, this result is analogous to the limits on the
chargino masses in the MSSM. For $M_{0}\gg  M_{Z}$ 
the off-diagonal terms become relatively unimportant and
 $\hat{\mu}^{2}\rightarrow (a-1) M_{2}^{2}.$ Therefore
the lightest chargino is mostly a wino with $m_{\chi^{+}_{1}}\simeq M_{2}$ 
and the other chargino has mass $m_{\chi^{+}_{2}}\simeq \sqrt{a-1} 
m_{\chi^{+}_{1}} > \sqrt{3} m_{\chi^{+}_{1}} . $

We next consider the neutralino sector, with five 
Majorana fermions. It is particularly important since it includes the 
lightest odd R-parity state, or LSP, and since its phenomenology can 
deviate from that predicted in the MSSM because of the presence 
of the singlet fermion (singlino,\ $\widetilde{s}).$\  The mass terms 
for the neutralinos are,

\begin{eqnarray}
&&\widetilde{s} \left[ 2\nu \widetilde{s}+\frac{%
\lambda }{\overline{g}}M_{Z}\left( \widetilde{h}_{1}\sin \beta +%
\widetilde{h}_{2}\cos \beta \right) \right]+\hat{\mu}\widetilde{h}_{1}\widetilde{h}_{2}\nonumber 
+M_{1}\widetilde{B}\widetilde{B}+M_{2}\widetilde{W}_{3}\widetilde{W}_{3}\\
&&+M_{Z}\left( \widetilde{B} \sin \theta _{\mathrm{w}}+
\widetilde{W}_{3}\cos \theta _{\mathrm{w}}\right) 
\left( \widetilde{h}_{1}\cos \beta +\widetilde{h}_{2}\sin \beta \right)
  \label{E6} 
\end{eqnarray}

The point to be stressed is the relatively small mixing of the singlino 
to the higgsinos. This is because $s\gg h_{2}>h_{1}$ and $(2/3)\left|
A_{0}\right| <\left| 12\nu \right| <\left| A_{0}\right| $ ,
$\hat{\mu}=\lambda s\gtrsim \sqrt{2}M_{0}>\sqrt{2M_{Z}}.$ Hence the
singlino remains an almost pure state of mass $\left|2 \nu \right|$, and
the higgsino-gaugino sector is analogous to the 
charged one. The experimental constraints on the chargino
imply, as just discussed, that the lightest chargino is
mostly a gaugino, which will also imply an analogous situation in the
non-singlet neutralino sector.

One finds indeed that the lightest states are \footnote{%
As a matter of fact, when the mass spectrum of the model is relatively low,
the mixings are more important in the neutralino sector, and, besides the
singlino, the lightest state can be more like a photino $(\widetilde{\gamma }
)$ than a ``bino'' $(\widetilde{B}).$ But this can be easily incorporated in
the discussion that follows where a $\widetilde{B}$-state is assumed for
simplicity.}: one which is mostly $\widetilde B,$ with a mass $\simeq
M_{1}\simeq M_{0}/\sqrt{6}$,  and another one, mostly
$\widetilde{W}_{3},$ with mass $  
\simeq M_{2}\simeq 2M_{1}.$ From the bound on the chargino mass one has 
$M_{1}^{2}>M_{Z}^{2}/5.$
The most important difference between the (M+1)SSM and the 
MSSM could be in the LSP character: unless one badly violates the
universality assumption for the scalar masses, the LSP is mostly a 
$\widetilde{B}$ in the MSSM, while the $\widetilde{s}$ is also a LSP 
candidate in the (M+1)SSM. Let us consider the condition for a 
{\sl singlino} LSP, 

\begin{equation}
\left| 2\nu \right| \lesssim M_{1}  \label{E7}
\end{equation}

\noindent which in terms of the bare parameters yields:

\begin{equation}
\left| A_{0}\right| < M_{0}\frac{\sqrt{6}}{3+\sqrt{1+8\alpha }}%
< \sqrt{\frac{3}{8}}M_{0}  \label{E8}
\end{equation}

This corresponds to some constraints in the coupling constants as well.\
From (\ref{A3}) one infers $\kappa /\overline{g} \lesssim .01.$ 
The $S$ v.e.v. has to be relatively large because of (\ref{A7}), which
gives $s/\left( \sqrt{h_{1}^{2}+h_{2}^{2}}\right) >20.$ From (\ref{A6})
and (\ref{E8}) one derives the constraint, $\left| \hat{\mu}/\nu \right|
=\lambda/\kappa  \gtrsim 5.$
On the other hand, (\ref{E8}) also implies $M^2_0 > 24$ $m^2_0$.
In other words, the light singlino scenario corresponds to gaugino masses 
being the largest soft terms, a ``gaugino dominated scenario''. This implies
many correlations between sparticle masses.

If the singlino is the LSP, the second lightest neutralino being mostly a $%
\widetilde{B}$ should be quite long-lived. Indeed, its decay into $%
\widetilde{s}$ has to go through its mixing to higgsinos. Then the decay
$\widetilde{B}\rightarrow \widetilde{s}$ is proportional to  
$g_{1}^{2}\left( h_{1}^{2}+h_{2}^{2}\right) /s^{2}<g_{1}^{2}/400.$ 
Conversely, if (\ref{E8})
is not realized, then $\widetilde{s}$ will decay into the LSP, mostly $%
\widetilde{B},$ with a similar coupling ($\widetilde{s}$ would be 
produced in Higgs or higgsino decays).

Let us now turn to the slepton sector. The lowest lying states
are the $\nu$ sneutrinos $\widetilde{\nu }_{\mathrm{e}}$ and the
``right-handed'' 
sleptons, $\widetilde{\ell }_{R},\ell =\mathrm{e},\mu ,\tau $ \footnote{%
In a $\mathrm{GUT}$ context, the $\tau $ superpartners can become
considerably splitted and much lighter than the other sleptons, as a
consequence of renormalization between the Planck and $\mathrm{GUT}$ scales,
as noticed in \cite{BHS}. However in this paper we are assuming $%
h_{t}>h_{\tau }$ at $M_{\mathrm{Planck}}$ and we consistently neglect this
possibility.}. This follows from the assumption of universality for the soft
terms.\ The approximated expressions for the slepton masses as given 
in the Appendix are

\begin{eqnarray}
m_{\widetilde{\ell _{R}}}^{2} &=&m_{0}^{2}+ .15 M_{0}^{2}+M_{Z%
}^{2}\sin ^{2}\theta _{W}\left| \cos 2\beta \right|  \nonumber \\
m_{\widetilde{\nu }}^{2} &=&m_{0}^{2}+ .5 M_{0}^{2}-\frac{1}{2}M_{%
Z}^{2}\left| \cos 2\beta \right|  \label{E9}
\end{eqnarray}

Searches at LEP 1  put a limit of roughly $M_{Z}^{2}/4$ on
both of them. For the relatively large values of $\tan \beta $ that tend to
prevail in the (M+1)SSM $|\cos 2\beta| \approx 1.$ With the LEP 1 results
on the chargino masses, as pointed out above, one had a bound $%
2M_{0}^{2}\gtrsim M_{Z}^{2}$ and the bounds on $m_{\widetilde{\nu }}^{2}$ 
gave some useful information.\ With the LEP 1.5 results, the limit on $M_{0}$
implied by (\ref{E6}) is  enough to ensure $m_{\widetilde{\nu }}>
m_{\widetilde{\ell _{R}}}.$

A more interesting issue is provided by the interplay between chargino and
slepton searches at LEP 200. In order to compare their masses, we first
replace $M_0$ in (\ref{E9}) in terms of the lightest chargino mass in
(\ref{E4})  to obtain

\begin{equation}
m_{\widetilde{\ell_R}}^{2}\simeq 
\frac{a}{4(a-1)}m_{\chi ^{+}}^{2}+
m_{0}^{2}+\frac{M_{Z}^{2}}{4}\left( 1+\frac{3}{4a}+\frac{3}{2\sqrt{a} 
\tan \beta} \right) \gtrsim
m_{0}^{2}+\frac{2}{7}M_{Z}^{2}  \label{E11}
\end{equation}

\noindent which gives a qualitative explanation of figure (2) in the
next section. The difference between $m_{\widetilde{\ell_R}}$
and $m_{\chi ^{+}}/2$ is almost a measure of the parameter $m_{0}$
in the future searches for these particles at relatively high energies. 
As a matter of fact, this reverses the situation  at LEP 1, where 
the chargino was expected to be lighter than the selectron. Indeed 
by comparing (\ref{E9}) and (\ref{E4}), we find that the chargino 
is always lighter than the selectron if
$M_{2}^{2}<M_{Z}^{2}/2,$ or, equivalently, as 
$m_{\chi ^{+}}^{2}<M_{Z}^{2}/3=(52.5 \mathrm{ GeV} )^2 .$ This is 
quite well reproduced by our numerical results. 

In particular, in the light singlino scenario, the term $m_0^2$
can be almost neglected in (\ref{E11}) so that the selectron and chargino
masses are strongly correlated.

\section{Numerical analysis, experimental and theoretical constraints}
\label{sec:cuts}

In this section we describe the numerical procedure employed to
produce the figures (1) - (6).  
As mentioned above, the essential independent parameters of the model are 
(apart from the lepton and light quark Yukawa couplings and the known 
gauge couplings) three Yukawa couplings 
$h_t$, $\lambda$ and $\kappa$ and, under the universality hypothesis at 
$\Lambda_0 = M_{GUT}$, the three soft parameters $M_0$, $m_0$ and $A_0$. Since 
these are the only dimensionful parameters in the theory and the scale is 
finally set by the experimental value of $M_Z$, the true independent
parameters  
are the five dimensionless quantities $h_{t_0}$, $\lambda_0$, $\kappa_0$, 
$m_0 / M_0$ and $A_0 / M_0$. 

The numerical results  presented here have been obtained by scanning 
over $\sim 10^6$ points in this five dimensional 
parameter space. (Here we used a logarithmic measure, but we have
checked the results  
with different measures as well. A lot of scanning of particular regions
of the parameter space
has been carried out in order to verify the boundaries of the parameter
space and the mass ranges, but the results of these scannings are not
shown in the figures.) In each case 
we integrate the one loop renormalization group equations down to the 
electroweak scale $v=174$ {GeV}. To one loop accuracy it is sufficient
to use $v$ as the low energy scale, independently from the exact  
particle masses of $O(v)$. At the one loop level, one can also take advantage
of (\ref{obvious}) and compute the coefficients in these equations,
what saves a lot of computer time to be used for a better scanning. 

Actually, also two loop renormalization group equations have been 
considered in \cite{King2}. In this case, however, the  
low energy scale has to be defined more precisely: The decoupling of
particles with masses of $O(v)$ has to be properly taken into account;
these masses, in turn, are only known once the Higgs potential (which
depends on these masses via the radiative corrections) has been
minimized. As a consequence, the  
numerical procedure becomes much more envolved and allows only to study much 
less points in the parameter space. On the other hand, the corresponding 
numerical results hardly deviate from ours. Individual points in the 
parameter space may well lead to somewhat different particle masses, once two 
loop contributions have been taken into account. Allowed mass ranges and 
correlations, however, remain practically unchanged. We found it much
more important numerically to minimize the full one loop Coleman-Weinberg 
Higgs potential including the non-logarithmic contributions than to 
include two loop logarithms. 

Having obtained the parameters at low energy it is most 
convenient to check first the absence of slepton v.e.v.'s according 
to eq. (\ref{T}) at the scale $\sim A_e/h_e$.  
Next we minimize in each remaining case the effective potential (\ref{F}) including the Coleman-Weinberg radiative corrections (\ref{G}),(\ref{H}), 
numerically.  We test, whether the  minimum with 
$\left\langle S \right\rangle $, $\left\langle H_1\right\rangle $  
and $\left\langle H_2 \right\rangle \neq 0$ is the lowest one, and dismiss the 
corresponding set of parameters otherwise. We calculate $\tan \beta $ and 
discard the initial conditions that lead to  $\tan \beta >30$ (since we 
assume $h_t \gg h_b$ in the present work; larger values 
turn out to be somewhat disfavoured in the model).  

In the remaining cases we determine the overall scale of the dimensionful 
parameters by identifying $\left\langle H_1^2 \right\rangle + \left\langle 
H_2^2 \right\rangle $ with $2 M_Z^2 / (g_1^2+g_2^2)$, 
and compute the physical masses of all particles. Then we impose the following 
experimental constraints: For the top quark pole mass $m_{t}$ (where we take 
the leading $QCD$ corrections to the pole mass into account \cite{tar}) we 
require $168\ \mathrm{GeV} < m_{t} < 192\ \mathrm{GeV}$ \cite{PDG}.
Next we impose the LEP $1.5$ lower bound on the lightest chargino mass
$m_{\chi^+}$: $m_{\chi^+} > 65$ GeV \cite{Aleph}. As a result, nearly all
other experimental bounds on new particles turn out to be satisfied    
automatically, with some exceptions discussed in section \ref{sec:spectrum}. 
Indeed, though the experimental bounds on sleptons, stop, gluinos, as well as 
on the decays of the $Z$ into neutralinos and Higgs scalars are implemented
in our code, these conditions only eliminate a few marginal points in the
parameter space. This remarkable property also can be seen from the plots in 
the next section.

Let us briefly discuss the range of the  
bare parameters, which turns out to be consistent with our theoretical
and experimental  
constraints. Concerning the Yukawa couplings we first remark that, as in
the MSSM, the  
known range of the top quark mass together with the solution of the RGE
for $h_t$ as  
given in the appendix allows easily to obtain the allowed range for
$h_{t_0}$: $m_{t}> 168$ GeV leads to $h_{t_0}> 0.468$. The other  
two bare Yukawa couplings $\lambda_0$ and $\kappa_0$ turn out to be
fairly small, $2.7\cdot 10^{-3} \lesssim \lambda_0 \lesssim 0.32$ and 
 $1.1 \cdot 10^{-4} \lesssim \kappa_0 \lesssim 0.33$, and relatively
closely related, $0.04 \lesssim \kappa_0/\lambda_0 \lesssim 1.14$. 

As already remarked below eq. (\ref{E5}), the experimental  
lower limit on $m_{\chi^+}$ implies a lower limit on the soft susy
breaking parameter $M_0$.  
The numerical analysis is in good agreement with the approximate
analytic result and leads  
to $M_0 \gtrsim 90 \ GeV$. No strict upper limits (in the absence of
fine tuning constraints)  
on the soft susy breaking parameters have been obtained. The scalar mass
term $m_0^2$ turned  
out to be unconstrained by our analysis; we assumed, however, $m_0^2 >
0$. The numerical  
analysis also confirms the allowed range for the third susy breaking
parameter $A_0$ as a  
function of $M_0$ and $m_0$, eq. (28). The corresponding lower limit on
$A_0$ is a  
particularity of the (M+1)SSM, since it is required by the need to
destabilize the scalar  
potential in the singlet direction. The correlations between the
particle masses and the  
bare parameters can finally be obtained from the approximate analytic
solutions given in  
sect (6) and the appendix; in table 1 we present the precise numerical
parameters and masses  
for two particular generic cases: one with $A_0$ negative and a
non-singlet LSP, and one with $A_0$ positive and a singlet LSP.
\vskip 1truecm  
 
\section{Predictions for the particle spetrum}
\label{sec:spectrum}

In order to allow to study the correlations among the particle 
masses we choose to plot all masses against  the lightest 
chargino mass $m_{\chi^+}$. Each point in these plots corresponds to 
one of the $\sim 1.5 \cdot 10^3$ points in the parameter space, which satisfy 
all our theoretical and phenomenological constraints. 
The density of points in the plots is clearly not uniform;  
regions of low density (if not completely empty 
and hence forbidden), typically  
towards larger particle masses, correspond to regions where more and more fine 
tuning is required. We will not, however, employ quantitative ``fine tuning 
constraints'' in this paper in order to constrain the particle masses from 
above; a reader, who is interested in ``probable particle masses'', can obtain 
them by investigating the relative densities in our plots.
Let us first consider the gluino mass $M_3$, fig. 1. 
As already noted in eq. (\ref{E5}), given the present experimental bounds,
$m_{\chi^{+}}$ and $M_3$ turn out to be strongly correlated. 
Fig. 1 agrees well with the approximate analytic relation 
(\ref{E5}), and one sees that the lower experimental bound on
$m_{\chi^+}$ turns into a lower bound $M_3 \gtrsim 260$ {GeV} on the
gluino mass within our model. For large masses, the ratio between the 
lightest chargino and the gluino is close to $g_2^2/g_3^2$ as predited by
the chargino being a wino, but slightly smaller as the parameters in the 
chargino mass matrices are dynamically correlated (as discussed in section
\ref{sec:inos}). 

Next we turn to the slepton masses. 
The lightest charged ``right-handed'' slepton  
mass $m_{\widetilde {\ell_R}}$ is plotted against $m_{\chi^+}$ in fig.
2. We see that  
$m_{\chi^+} > 65$ {GeV} implies $m_{\widetilde \ell_R} > 55$ {GeV} in our 
model. In most cases, as noted below  (\ref{E11}), 
the charged slepton mass $m_{\widetilde \ell_R}$  has the tendency to 
be smaller than $m_{\chi^+}$ (except for large values of the  
bare scalar mass $m_0$), although it always satisfies 
$m_{\widetilde \ell_R} \gtrsim m_{\chi^+}/2$ in agreement with 
the analytic approximation (\ref{E11}).  In fig. 3 
we plot the sneutrino mass $m_{\tilde \nu}$ against $m_{\chi^+}$. Given 
$m_{\chi^+} > 65$ GeV, the {LEP 1} lower limit on $m_{\tilde \nu}$ only 
eliminates a tiny additional corner in the parameter space. 

The neutral scalar Higgs sector has already been discussed in detail in 
ref. \cite{ERS2}, in particular with respect to future experiments at
LEP $2$.  Although in \cite{ERS2} only the LEP $1$ bound on 
the chargino mass $m_{\chi^+}$  had been taken into account, 
the results concerning the neutral Higgs sector remain essentially 
unchanged. Therefore we briefly repeat only the essential features. 
In order to take the experimental constraints on neutral Higgs bosons  
properly into account both their masses $and$ their couplings 
to the $Z$ bosons   have to be evaluated. 
In fact, the negative LEP $1$ results on neutral CP-even Higgs scalars 
\cite{PDG} eliminate only a tiny region of the parameter space of the  
present model, corresponding to $m_h < 58$ GeV and a coupling $hZZ$ as
large as  
in the non-supersymmetric standard model. On the other hand, a large
region with  
smaller values of $m_h$, but small couplings $hZZ$ remains unconstrained 
by LEP $1$.  
In this region the lightest CP-even Higgs scalar is essentially a gauge
singlet  
and hence decoupled. In view of this possibility the experimental searches for 
and the theoretical upper limits on the masses of the lightest CP-even Higgs 
scalar have to be turned into considerations of the ``lightest visible CP-even 
Higgs scalar''. Fortunately it turns out within the present model
\cite{Ellis,King,Kam}, that upper limits exist also on the lightest
visible CP-even 
Higgs scalar (which could be the second lightest), varying from $140$ to
$160$ GeV for gluino masses (as a measure of the susy breaking scale)
from $1$ to $3$ TeV. Similar complications arise also in the CP-odd
Higgs sector, which contains a gauge singlet state as well.  
Here no upper limit can be derived, and we just remark that a visible CP-odd 
Higgs scalar below $130$ GeV is impossible within the present model and the 
present experimental constraints on $m_{\chi^+}$.

As in the MSSM the mass of the visible (non-singlet) CP-odd Higgs scalar
is actually just somewhat below the mass $m_{H^+}$ of the charged Higgs;
cf. the upper left entry  
of the CP-odd mass matrix eq. (15) and eq. (16). In fig. 4 we plot
$m_{H^+}$ against  
$m_{\chi^+}$, and we see that the present lower limit on $m_{\chi^+}$ implies 
$m_{H^+} > 160\ GeV$ (increasing rapidly with increasing lower limits on 
$m_{\chi^+}$). 

The range of $\tan{\beta} $ is fairly restricted, and depends strongly
on the sign of the bare susy breaking parameter $A_0$: for $A_0 > 0$ we
have $\tan{\beta} < -2.6$, whereas for $A_0 < 0$ we find both signs,
but large absolute values for $\tan{\beta}$: $\tan{\beta} < -6.5$ or 
$\tan{\beta} > 8.7$. Since these inequalities hardly depend on the mass
of the chargino, a figure does not provide additional information. 

Next we turn to the squarks. Because of renormalization effects between
$M_{GUT}$ and $M_Z$ (cf. the appendix) the lightest stop eigenstate is
the lightest of all squarks; subsequently we concentrate on this
particle. In fig. 5 we plot its mass $m_{T_1}$ against $m_{\chi^+}$, and
we see that the present lower limit on $m_{\chi^+}$ implies $m_{T_1} >
170$ GeV. A possible ``supersymmetric'' top quark decay mode is thus
practically excluded within the present  model. 

Finally we consider the neutralino sector. As discussed in chapter (6),
two different  
scenarios are possible within the present model, depending on the mass
of the nearly  
pure singlino: In the case of a heavy singlino, the two lightest
neutralino states are  
a practically pure bino with a mass $M_1$ and a practically pure wino
with a mass  
$M_2 \sim 2 M_1$. The numerical procedure confirms this approximate
analytic result to  
a very high accuracy. Since furthermore the chargino mass $m_{\chi^+}$
is close to $M_2$, cf. eq. (\ref{E5}), the two lightest neutralino
masses can easily  
be obtained as a function of $m_{\chi^+}$ and a figure is of no further
use. Here the  
LSP is, of course, the bino, and the supersymmetric decays of all
sparticles will  
proceed as in the MSSM with a suitable set of parameters. 

The case of a light singlino  
is actually more interesting. From the numerical analysis we find, that
now the second  
and third neutralinos correspond to the above mentioned nearly pure bino
and wino states.  
In fig. 6 we plot the mass of the lightest singlino, $m_{ls}$, against
the bino mass  
$M_1$ (for those parameters where $m_{ls} < M_1$). Now supersymmetric
decays of all  
sparticles will start as in the MSSM, until the bino state is
produced. (A direct decay  
into the singlino LSP will be heavily suppressed due to its tiny
coupling.) Finally the  
bino will decay into the singlino LSP, producing an additional cascade,
whose consequences  
on experimental susy searches still have to be worked out in more detail.

\vskip 1truecm

\section{Conclusions and outlook}
\label{sec:conclusions}

In the present paper we have presented an exhaustive overview over the
parameter space  
and the particle spectra, which are consistent with the correct vacuum
structure and  
present experimental constraints, within the (M+1)SSM under the
universality hypothesis.  
Since the scalar potential and the corresponding minimization conditions
differ considerably  
 from the MSSM due to the presence of the additional singlet
superfield, the outcome of  
this analysis was far from predictable. We have seen that still a large
region in the  
parameter space consistent with all constraints exists. This region is
characterized,  
however, by a) small values of the new Yukawa couplings $\lambda$ and
$\kappa$, b) a large  
value of the scalar singlet $vev$, c) a strong tendency of the singlet
states to decouple  
 from the rest of the spectrum. In spite of the complexity of the model
this region allows  
for fairly accurate analytical approximations to the vacuum minimization
conditions and  
particle masses, which we have worked out and presented in some detail. 

The near decoupling of the singlet states leads to a particle spectrum,
which is close to  
the MSSM for a suitable set of the corresponding parameters. Our
figures show strong  
correlations or inequalities among the particle masses, which allow for
experimental  
falsifications of our underlying hypothesises. For instance, as in the
MSSM with the  
universality hypothesis, sparticles are not light enough to play an
important role for  
the radiative corrections to $R_{b \bar b}$ \cite{rbb}. Only the Higgs
boson, the lightest chargino,  
sleptons and neutralinos could be light enough to be discovered at LEP
2; gluinos,  
squarks and charged or CP-odd Higgs scalars are necessarily too heavy in
our model.  

Our approximate analytic results allow to translate future possibly
negative sparticle  
searches into constraints on the soft susy breaking parameters. Obvious
are lower limits  
on the universal gaugino mass $M_0$ from negative chargino searches (cf.
eq. (\ref{E5})) and,  
of course, negative gluino searches. From eqs. (50), one can furthermore
easily deduce  
excluded regions in the $m_0 - M_0$ plane from negative ``right-handed''
slepton and  
sneutrino searches. These relations are actually not different from the
MSSM. Eq. (28),  
however, allows to turn these constraints on $m_0$ and $M_0$ into
constraints on $A_0$,  
which is a particular feature of the (M+1)SSM. 

If future sparticle searches turn out to be successful, and the mass
pattern predicted by  
our model turns out to be confirmed, the question arises whether the
model can be  
distinguished from the MSSM with corresponding parameters. The answer
depends on the  
corresponding parameters of the (M+1)SSM. First, there still exist
small corners in the  
parameter space, where mixing with the singlet states either in the
neutral Higgs sector  
or in the neutralino sector could be possible; in this case the
couplings of the  
corresponding visible particles could be substantially and hence
measurably smaller than  
in the MSSM. A more likely possibility - but by no means certain - is
the singlino LSP scenario. As  
discussed in the previous section, in this case sparticle decays will
produce an additional  
cascade with respect to the MSSM. This scenario desserves further
phenomenological investigations.
\newpage

\begin{appendix}
\section{Appendix}

In this Appendix we list the analytic solutions of the RGE for soft susy
breaking terms in the approximation where the dependence on all Yukawa
couplings but the top one, $h_{t},$ are neglected.

Also, flavour mixing is neglected in these expressions (see \cite{Brasa})
for solutions that fully take family mixing into account). However, for the
sake of generality, universality is {\it not} assumed for the soft terms
at the unification scale, $\Lambda _{0}.$ All quantities are defined at the
scale $\Lambda $ (unless stated otherwise) and the boundary conditions at
the scale $\Lambda _{0},$ are generally denoted as $X_{0}=X(\Lambda _{0}).$
We define the following quantities:

\begin{eqnarray}
t &=&\frac{1}{16\pi ^{2}}\ln \frac{\Lambda _{0}}{\Lambda }  \nonumber \\
\Pi _{(\mathrm{i})} &=&\prod\limits_{\alpha =1}^{3}\left( \frac{g_{\alpha
}^{2}}{g_{\alpha _{0}}^{2}}\right) ^{\frac{2C_{\alpha }^{(\mathrm{i})}}{%
b_{\alpha }}}  \nonumber \\
J &=&\Pi ^{-1}_{(u)}(t)\int_{0}^{t}\Pi _{(u)}(t^{\prime })dt^{\prime } 
  \label{Z1}
\end{eqnarray}

\noindent The index $\alpha =1,2,3,$ refers to the $U(1), SU(2)$ and
$ SU(3)$ gauge groups, respectively, and $C_{\alpha }^{(\mathrm{i})}$
is the sum of the their quadratic Casimir eigenvalues for the three fields
in the i-th Yukawa coupling. We replace the integral $J $ by 
the value $h_{\mathrm{crit}}$ of $h_{t}$ at the scale $\Lambda $ 
corresponding to a Landau pole at $ \Lambda _{0},$ by defining

\begin{eqnarray}
\frac{1}{h_{\mathrm{crit}}^{2}} =12J\qquad \qquad \quad \rho =\frac{%
h_{t}^{2}}{h_{\mathrm{crit}}^{2}}      \label{Z2}
\end{eqnarray}

The solutions for the Yukawa coupling are then as follows

\begin{eqnarray}
\lambda ^{2} &=&\lambda _{0}^{2}\Pi _{(\lambda )}\left( 1-\rho \right) ^{1/2}
\nonumber \\
h_{t}^{2} &=&h_{t_{0}}^{2}\Pi _{(u)}\left( 1-\rho \right)   \nonumber \\
h_{c(u)}^{2} &=&h_{c(u)_{0}}^{2}\Pi ^{(u)}\left( 1-\rho \right) ^{1/2} 
\nonumber \\
h_{b}^{2} &=&h_{b_{0}}^{2}\Pi _{(d)}\left( 1-\rho \right) ^{1/6}  \nonumber
\\
h_{s(d)}^{2} &=&h_{s(d)_{0}}^{2}\Pi _{(d)}  \nonumber \\
h_{\ell }^{2} &=&h_{\ell _{0}}^{2}\Pi _{(\ell )}\quad \left( \ell =\mathrm{e}%
,\mu ,\tau \right)   \nonumber \\
\kappa ^{2} &=&\kappa _{0}^{2}    \label{Z3}
\end{eqnarray}

The soft trilinear terms are as follows:

\begin{eqnarray}
A_{\lambda } &=&A_{\lambda_0 }-\frac{1}{2}\rho \left( A_{t_{0}}+\xi
M_{0}\right) +2M_{0}t\left( 3g_{2}^{2}+g_{1}^{2}\right)   \nonumber \\
A_{\kappa } &=&A_{\kappa 0}  \nonumber \\
A_{t} &=&A_{t_{0}}-\rho \left( A_{t_{0}}+\xi M_{0}\right)+2M_{0}t\left( 
\frac{16}{3}g_{3}^{2}+3g_{2}^{2}+\frac{13}{9}g_{1}^{2}\right)   \nonumber \\
A_{c(u)} &=&A_{c(u)0}-\frac{1}{2}\rho \left( A_{t_{0}}+\xi M_{0}\right)
+2M_{0}t\left( \frac{16}{3}g_{3}^{2}+3g_{2}^{2}+\frac{13}{9}g_{1}^{2}\right) 
\nonumber \\
A_{b} &=&A_{b0}-\frac{1}{6}\rho \left( A_{t_{0}}+\xi M_{0}\right)+2M_{0}t
\left( \frac{16}{3}g_{3}^{2}+3g_{2}^{2}+\frac{1}{9}g_{1}^{2}\right)  
\nonumber \\
A_{s(d)} &=&A_{s(d)0}+2M_{0}t\left( \frac{16}{3}g_{3}^{2}+3g_{2}^{2}+\frac{1}{%
9}g_{1}^{2}\right)   \nonumber \\
A_{\tau (\mu ,e)} &=&A_{\tau (\mu ,e)0}+2M_{0}t\left( 
3g_{2}^{2}+3g_{1}^{2}\right)   \nonumber \\
\xi  &=&\left( \frac{t}{J}-1\right) =\left( 12h_{\mathrm{crit}}^{2}t-1\right) 
  \label{Z4}
\end{eqnarray}

Soft supersymmetry breaking scalar masses can also be written in terms 
of $\rho ,t$ and J (or $\xi ),$ as follows:

\begin{eqnarray}
m_{S}^{2} &=&m_{S0}^{2}  \nonumber \\
m_{1}^{2} &=&m_{10}^{2}+g_{1}^{2}ts_{0}+2\gamma _{L}tM_{0}^{2}  \nonumber \\
m_{2}^{2} &=&m_{20}^{2}-\frac{3\rho }{2}\overline{m}_{0}^{2}-\frac{\rho }{2}%
K-g_{1}^{2}ts_{0}+2\gamma _{L}tM_{0}^{2}  \nonumber \\
m_{T}^{2} &=&m_{T0}^{2}-\rho \overline{m}_{0}^{2}-\frac{\rho }{3}K+\frac{4}{3%
}g_{1}^{2}ts_{0}+2\gamma _{U}tM_{0}^{2}  \nonumber \\
m_{Q_{3}}^{2} &=&m_{Q30}^{2}-\frac{\rho }{2}\overline{m}_{0}^{2}-\frac{\rho 
}{6}K-\frac{1}{3}g_{1}^{2}ts_{0}+2\gamma _{Q}tM_{0}^{2}  \nonumber \\
m_{B}^{2} &=&m_{B0}^{2}-\frac{2}{3}g_{1}^{2}ts_{0}+2\gamma _{D}tM_{0}^{2} 
\nonumber \\
m_{U_{i}}^{2} &=&m_{Ui0}^{2}+\frac{4}{3}g_{1}^{2}ts_{0}+2\gamma _{U}tM_{0}^{2}
\nonumber \\
m_{Q_{i}}^{2} &=&m_{Qi0}^{2}-\frac{1}{3}g_{1}^{2}ts_{0}+2\gamma _{Q}tM_{0}^{2}
\nonumber \\
m_{D_{i}}^{2} &=&m_{Di0}^{2}-\frac{2}{3}g_{1}^{2}ts_{0}+2\gamma _{D}tM_{0}^{2}
\nonumber \\
m_{Lj}^{2} &=&m_{Lj0}^{2}+g_{1}^{2}ts_{0}+2\gamma _{L}tM_{0}^{2}  \nonumber \\
m_{Ej}^{2} &=&m_{Ej0}^{2}-2g_{1}^{2}ts_{0}+2\gamma _{E}tM_{0}^{2}  \nonumber \\
\gamma _{L} &=&3g_{2}^{2}\left( 1-g_{2}^{2}t\right) +
g_{1}^{2}\left( 1-11g_{1}^{2}t\right)   \nonumber \\
\gamma _{U} &=&\frac{16}{3}g_{2}^{2}\left( 1+3g_{3}^{2}t\right) +\frac{16}{9}%
g_{1}^{2}\left( 1-11g_{1}^{2}t\right)   \nonumber \\
\gamma _{Q} &=&\frac{16}{3}g_{3}^{2}\left( 1+3g_{3}^{2}t\right) +3
g_{2}^{2}\left( 1-g_{2}^{2}t\right) +\frac{1}{9}g_{1}^{2}\left(
1-11g_{1}^{2}t\right)   \nonumber \\
\gamma _{D} &=&\frac{16}{3}g_{3}^{2}\left( 1+3g_{3}^{2}t\right) +\frac{4}{9}%
g_{1}^{2}\left( 1-11g_{1}^{2}t\right)   \nonumber \\
\gamma _{E} &=&4g_{1}^{2}\left( 1-11g_{1}^{2}t\right)   \nonumber \\
s_{0} &=&\left[ m_{H_{2}}^{2}-m_{H_{1}}^{2}+\sum _i\left(
m_{Q_i}^{2}+m_{D_i}^{2}+m_{E_i}^{2}-m_{L_i}^{2}
-2m_{U_i}^{2}\right) \right] _{t=0} 
\nonumber \\
\overline{m}_{0}^{2} &=&\frac{1}{3}\left(
m_{2}^{2}+m_{T}^{2}+m_{Q}^{2}\right) _{t=0}  \nonumber \\
K  &=&\left( 1-\rho \right) \left( A_{t_{0}}+\xi M_{0}\right) ^{2}-\xi
^{2}M_{0}^{2}+  \nonumber \\
&&+2\left( \frac{16}{3}g_{3}^{2}+3g_{2}^{2}+\frac{13}{9}g_{1}^{2}\right)
\left( \xi +1\right) tM_{0}^{2}    \label{Z5}
\end{eqnarray}

Finally, let us recall that the renormalization effects on the 
effective variables $\hat{B}$ and $\hat{\mu }$ are easily obtained
by taking the limit in which the singlet fields are decoupled
at its classical value. Then,

\begin{eqnarray}
\hat{B} &=& A_{\lambda }+ \nu =\hat{B}_{0}-\frac{1}{2}\rho \left( A_{t_{0}}+\xi
M_{0}\right) +2M_{0}t\left( 3g_{2}^{2}+g_{1}^{2}\right)   \nonumber \\
\hat{\mu } ^2 &=& \lambda ^2 s^2 = \hat{\mu }_{0}^{2} \Pi ^{(\lambda )}
\left( 1-\rho \right) ^{1/2}   \label{Z6}
\end{eqnarray}
\end{appendix}



\vsize=25 truecm 
\parindent=12pt
\baselineskip=12pt
\overfullrule=0pt

\section*{Figure captions}
\newcounter{fig}
\begin{list}{\bf Figure \arabic{fig}:}{\usecounter{fig}}

\item Gluino mass $M_3$ versus the chargino mass $m_{\chi^+}$ (all in GeV)
\item Slepton mass $m_{\widetilde \ell_R}$ versus $m_{\chi^+}$
\item Sneutrino mass $m_{\widetilde \nu}$ 
versus $m_{\chi^+}$
\item Lightest stop mass $m_{T_1}$ 
versus $m_{\chi^+}$
\item Charged Higgs mass $m_{H^+}$ versus $m_{\chi^+}$
\item Mass of the second lightest neutralino $m_{\chi^0_2}$ versus the
mass of the  lightest neutralino $m_{\chi^0_1}$ in those cases where
$\chi^0_1$ is essentially a singlino  
\end{list}

\section*{Table caption}
\newcounter{tab}
\begin{list}{\bf Table \arabic{tab}:}{\usecounter{tab}}
\item Parameters and particle masses (in GeV) for two particular cases
(with $M_{susy}$ not too large). The three neutral CP even Higgs scalars
are denoted by $h_1$, $h_2$ and $h_3$ (in the order of increasing
masses), the two neutral CP odd Higgs scalars by $p_1$ and $p_2$, and
the two lightest neutralinos by $\chi^0_1$ and $\chi^0_2$, respectively.
In the first case $\chi^0_1$ is essentially a non-singlet, in the second
case it practically a pure singlino. 
\end{list}
\newpage
\pagestyle{empty}
\baselineskip=7pt
\centerline{\bf Table 1}
\vskip 5 mm
$$\vbox{\offinterlineskip \halign{
\tv#  &\cc{#} &\tv# &\cc{#} &\tv# &\cc{#} &\tv#  \cr
\noalign{\hrule}
&\cc{$\lambda_0$} &&\cc{$6,89\cdot 10^{-3}$} &&\cc{$2.68\cdot 10^{-3}$}  & \cr 
\noalign{\hrule}
&$\kappa_0$ &&$2.16\cdot 10^{-3}$ &&$1.08\cdot 10^{-4}$  & \cr
\noalign{\hrule}
&$h_{t_0}$ &&0.470 &&0.534 & \cr
\noalign{\hrule}
&$M_0$ &&104 &&116  & \cr
\noalign{\hrule}
&$A_0$ &&- 95.8 &&11.6  & \cr
\noalign{\hrule}
&$m_0$ &&22 &&0.6  & \cr
\noalign{\hrule}
&$\lambda$ &&$7,18\cdot 10^{-3}$ &&$2.67\cdot 10^{-3}$  & \cr
\noalign{\hrule}
&$\kappa$ &&$2.16\cdot 10^{-3}$ &&$1.08\cdot 10^{-4}$  & \cr
\noalign{\hrule}
&$h_t$ &&0.932 &&0.966  & \cr
\noalign{\hrule}
&$A_{\lambda}$ &&- 46.7 &&31.8  & \cr
\noalign{\hrule}
&$A_{\kappa}$ &&- 95.8 &&11.6  & \cr
\noalign{\hrule}
&$A_t$ &&- 273 &&- 256  & \cr
\noalign{\hrule}
&$s$ &&19500 &&- 53300  & \cr
\noalign{\hrule}
&$|\tan (\beta )|$ &&8.17 &&16.0  & \cr
\noalign{\hrule}
&$m_{top}$ &&170 &&177  & \cr
\noalign{\hrule}
&$M_3$ &&280 &&313  & \cr
\noalign{\hrule}
&$m_{h_1}$ &&55.1 &&8.00  & \cr
\noalign{\hrule}
&singlet component of $h_1$ &&.999 &&1.00  & \cr
\noalign{\hrule}
&$m_{h_2}$ &&101 &&108 & \cr
\noalign{\hrule}
&$m_{h_3}$ &&156 &&157  & \cr
\noalign{\hrule}
&$m_{p_1}$ &&110 &&14.2  & \cr
\noalign{\hrule}
&$m_{p_2}$ &&152 &&156  & \cr
\noalign{\hrule}
&$m_{H^+}$ &&172 &&176  & \cr
\noalign{\hrule}
&$m_{\tilde{\ell_R}}$ &&63.2 &&62.9  & \cr
\noalign{\hrule}
&$m_{\tilde{\nu}}$ &&46.5 &&54.8  & \cr
\noalign{\hrule}
&$m_{\chi^+}$ &&72.9 &&65.9 & \cr
\noalign{\hrule}
&$m_{T_1}$ &&179 &&214 & \cr
\noalign{\hrule}
&$m_{\chi^0_1}$ &&42.0 &&11.5 & \cr
\noalign{\hrule}
&singlet component of $\chi^0_1$ &&0.01 &&1.00 & \cr
\noalign{\hrule}
&$m_{\chi^0_2}$ &&71.4 &&39.1 & \cr
\noalign{\hrule}
}}$$

\begin{figure}[p]
\unitlength1cm
\begin{picture}(12,20)
\put(-1.8,20.5){$\bf M_3$}
\put(16.0,1.5){$\bf m_{\chi^+}$}
\put(6.0,0.5){\bf Figure 1}
\put(-1.9,-4){\epsffile{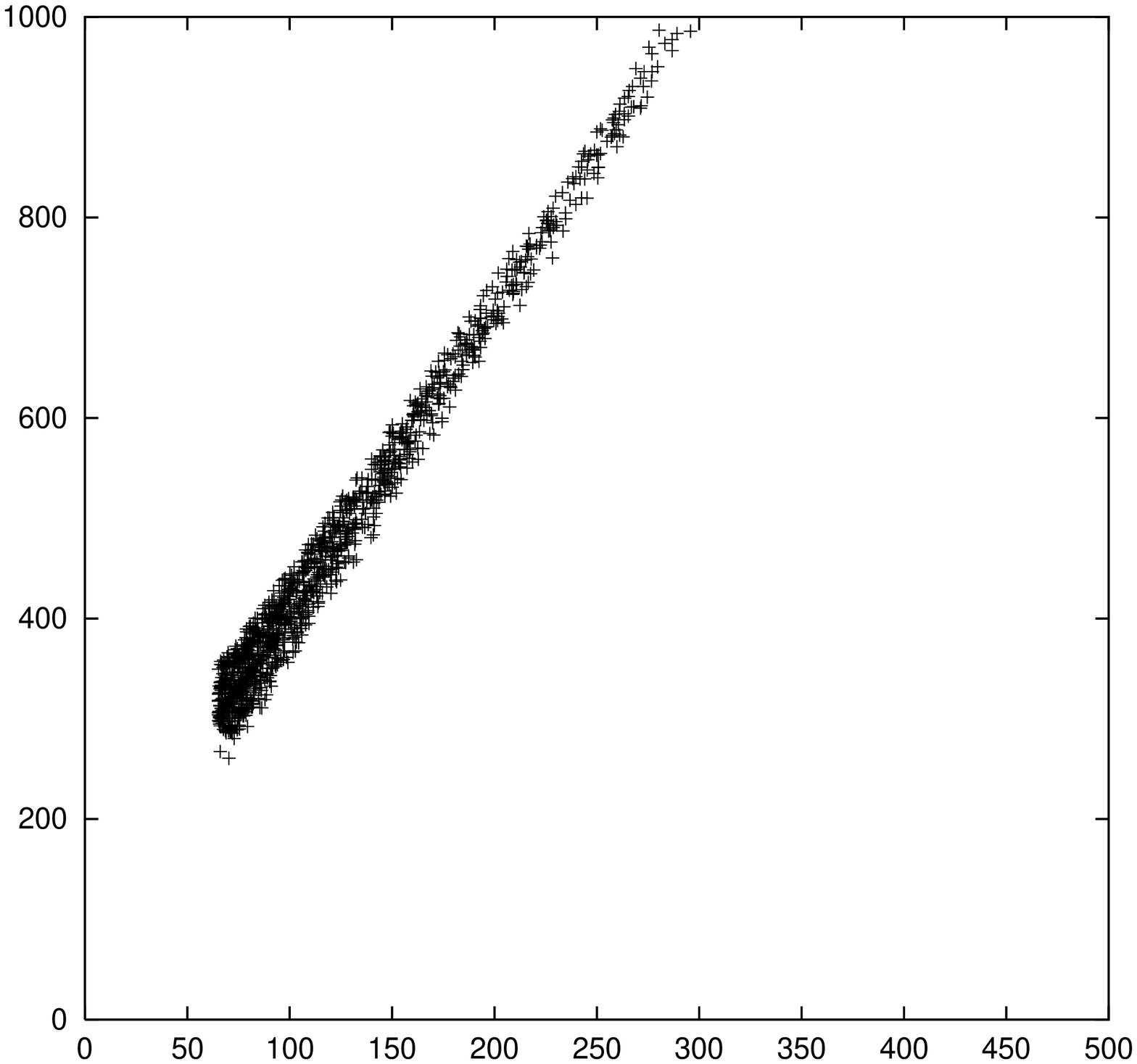}}
\end{picture}
\end{figure}

\begin{figure}[p]
\unitlength1cm
\begin{picture}(12,20)
\put(-1.8,20.5){$\bf m_{\widetilde \ell_R}$}
\put(16.0,1.5){$\bf m_{\chi^+}$}
\put(6.0,0.5){\bf Figure 2}
\put(-1.9,-4){\epsffile{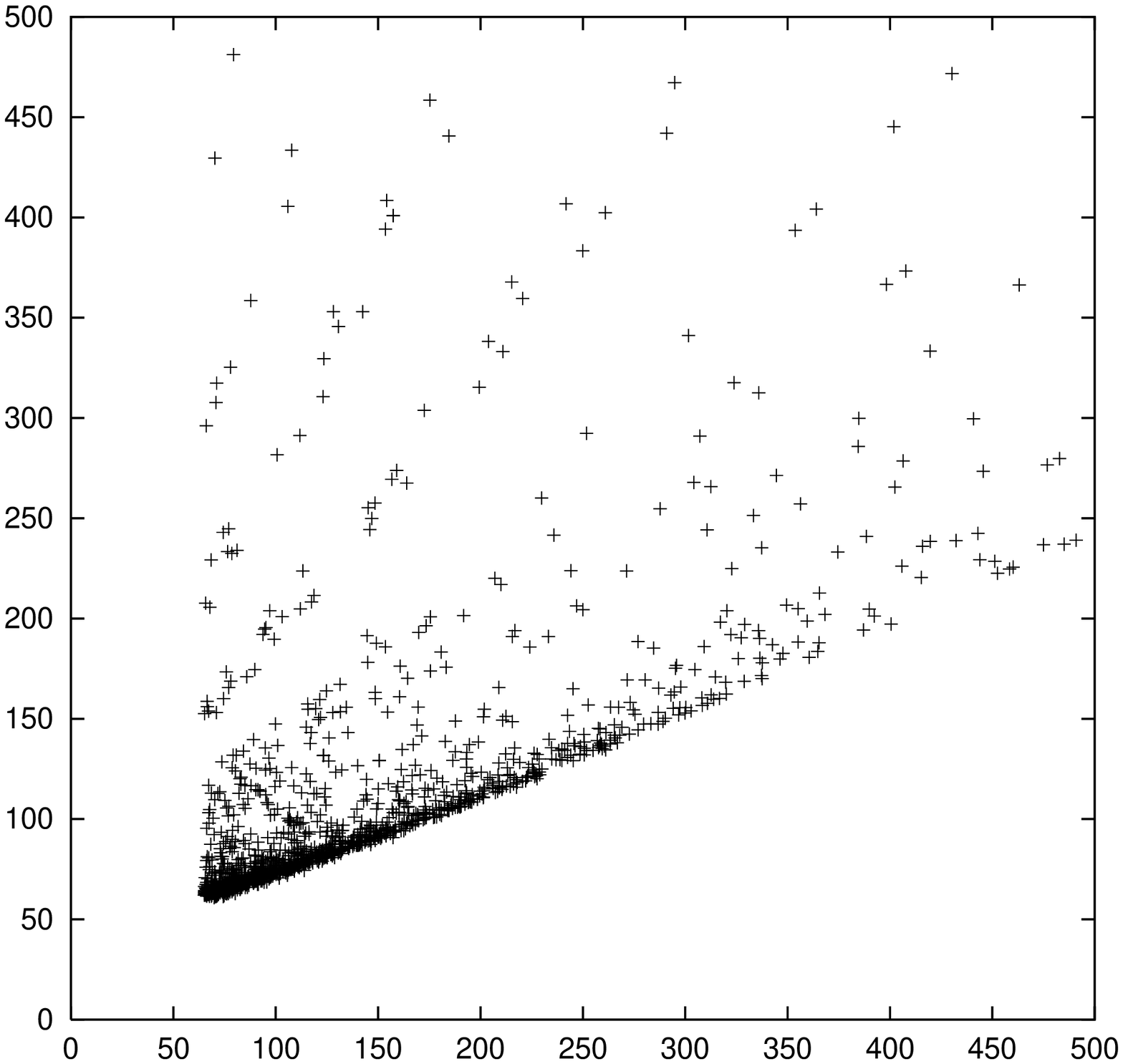}}
\end{picture}
\end{figure}

\begin{figure}[p]
\unitlength1cm
\begin{picture}(12,20)
\put(-1.8,20.5){$\bf m_{\widetilde \nu}$}
\put(16.0,1.5){$\bf m_{\chi^+}$}
\put(6.0,0.5){\bf Figure 3}
\put(-1.9,-4){\epsffile{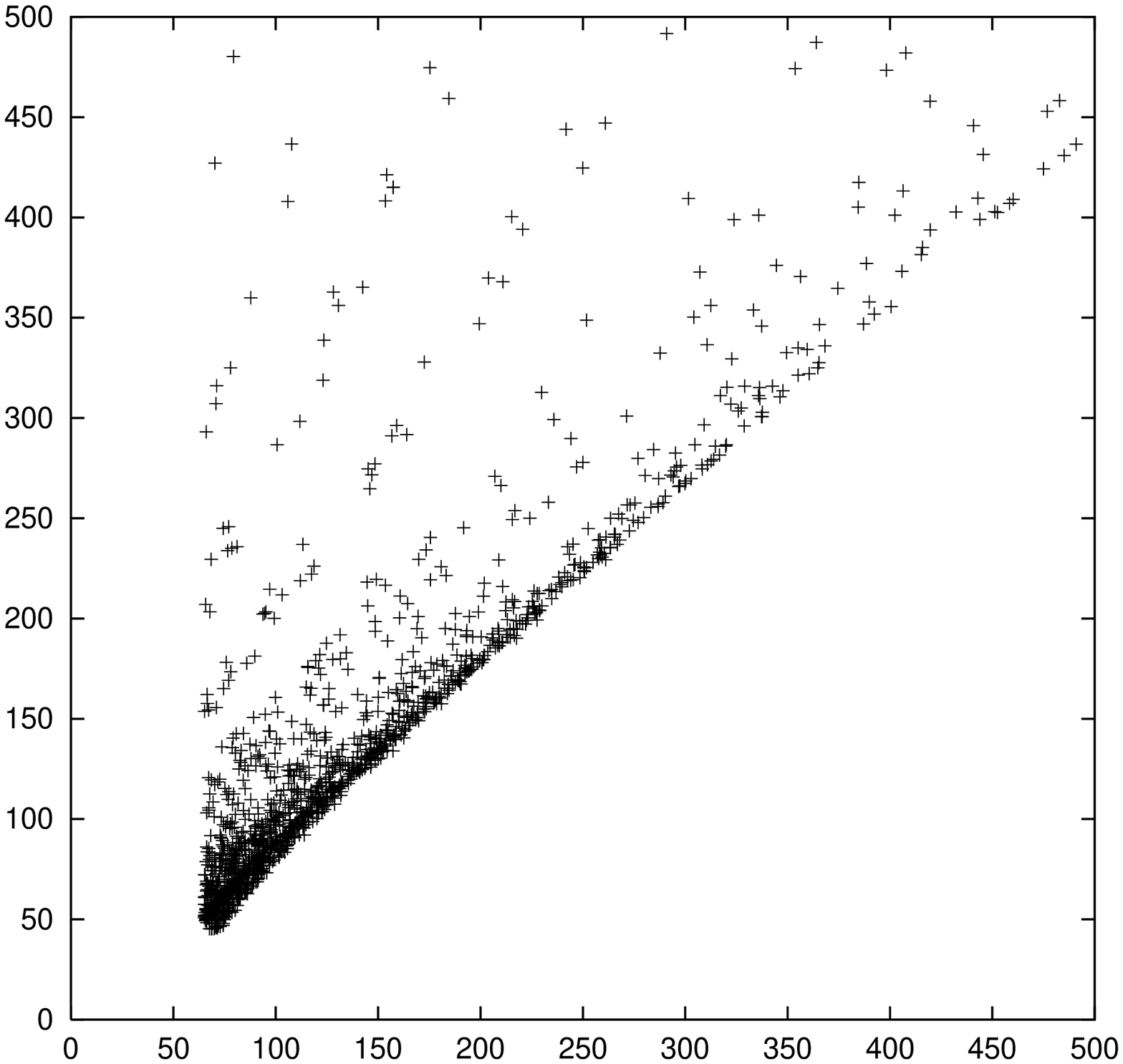}}
\end{picture}
\end{figure}

\begin{figure}[p]
\unitlength1cm
\begin{picture}(12,20)
\put(-1.8,20.5){$\bf m_{T_1}$}
\put(16.0,1.5){$\bf m_{\chi^+}$}
\put(6.0,0.5){\bf Figure 4}
\put(-1.9,-4){\epsffile{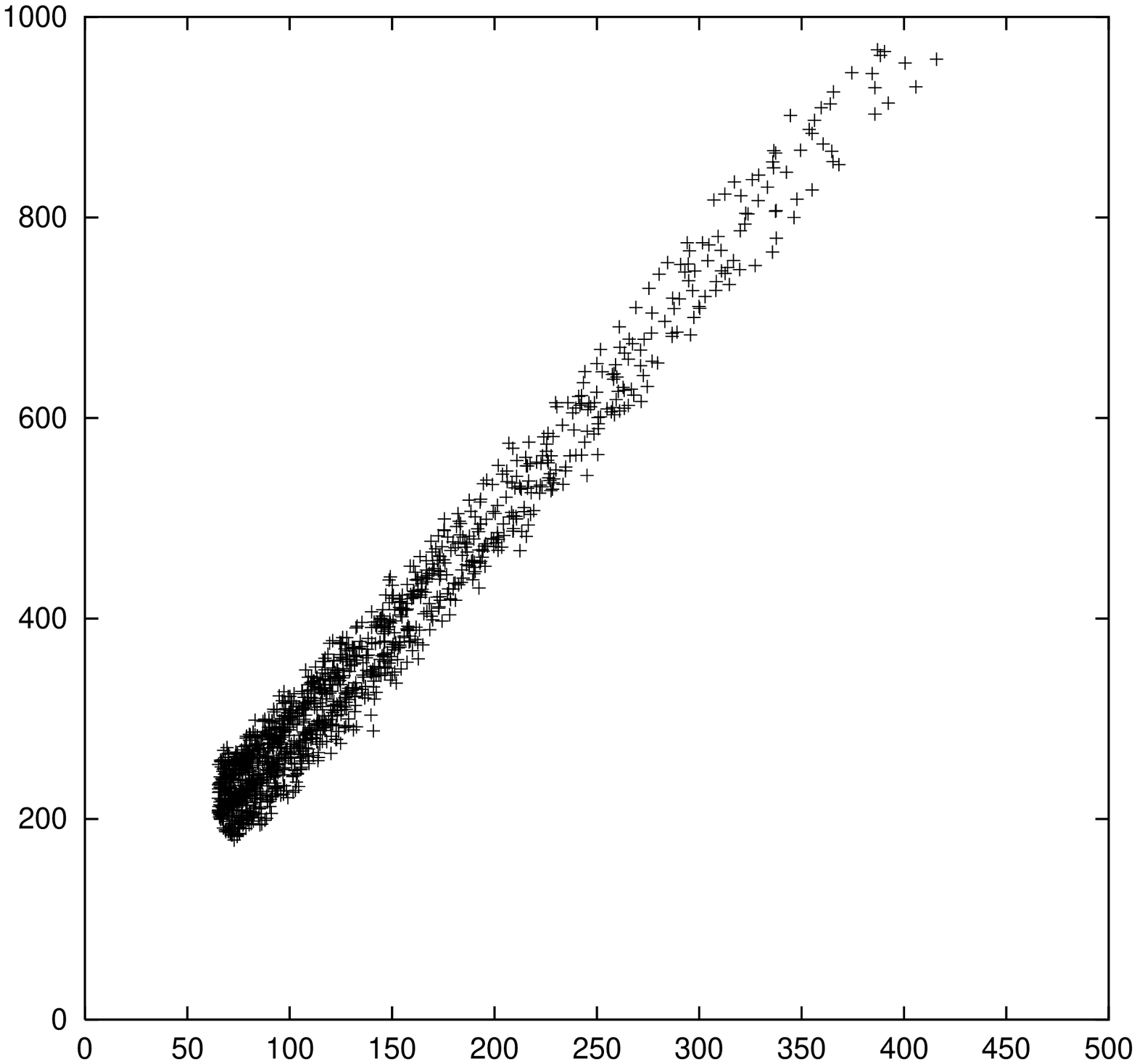}}
\end{picture}
\end{figure}

\begin{figure}[p]
\unitlength1cm
\begin{picture}(12,20)
\put(-1.8,20.5){$\bf m_{H^+}$}
\put(16.0,1.5){$\bf m_{\chi^+}$}
\put(6.0,0.5){\bf Figure 5}
\put(-1.9,-4){\epsffile{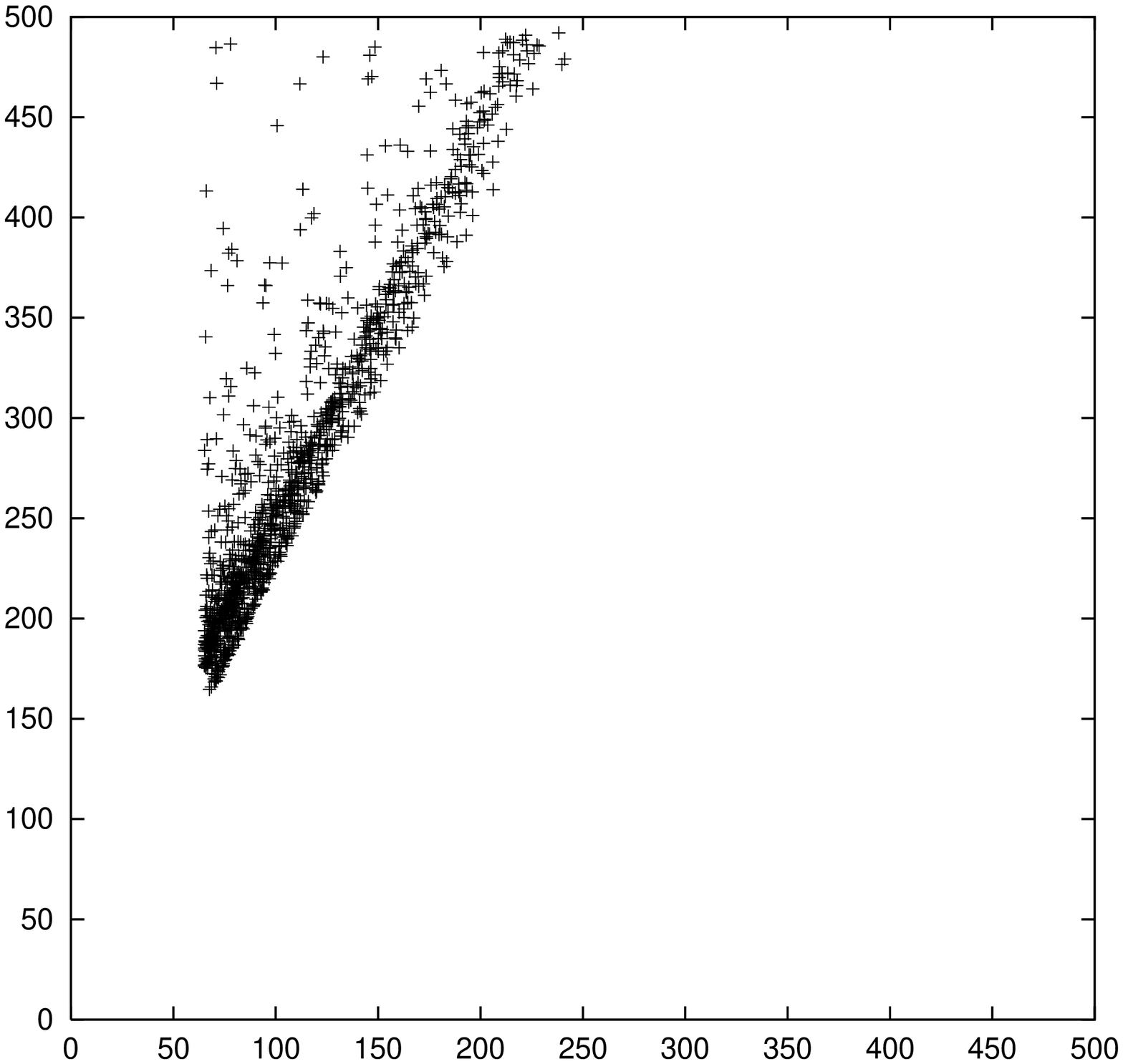}}
\end{picture}
\end{figure}

\begin{figure}[p]
\unitlength1cm
\begin{picture}(12,20)
\put(-1.8,20.5){${\bf m_{\chi^0_2}}$}
\put(16.0,1.5){${\bf m_{\chi^0_1}}$}
\put(6.0,0.5){\bf Figure 6}
\put(-1.9,-4){\epsffile{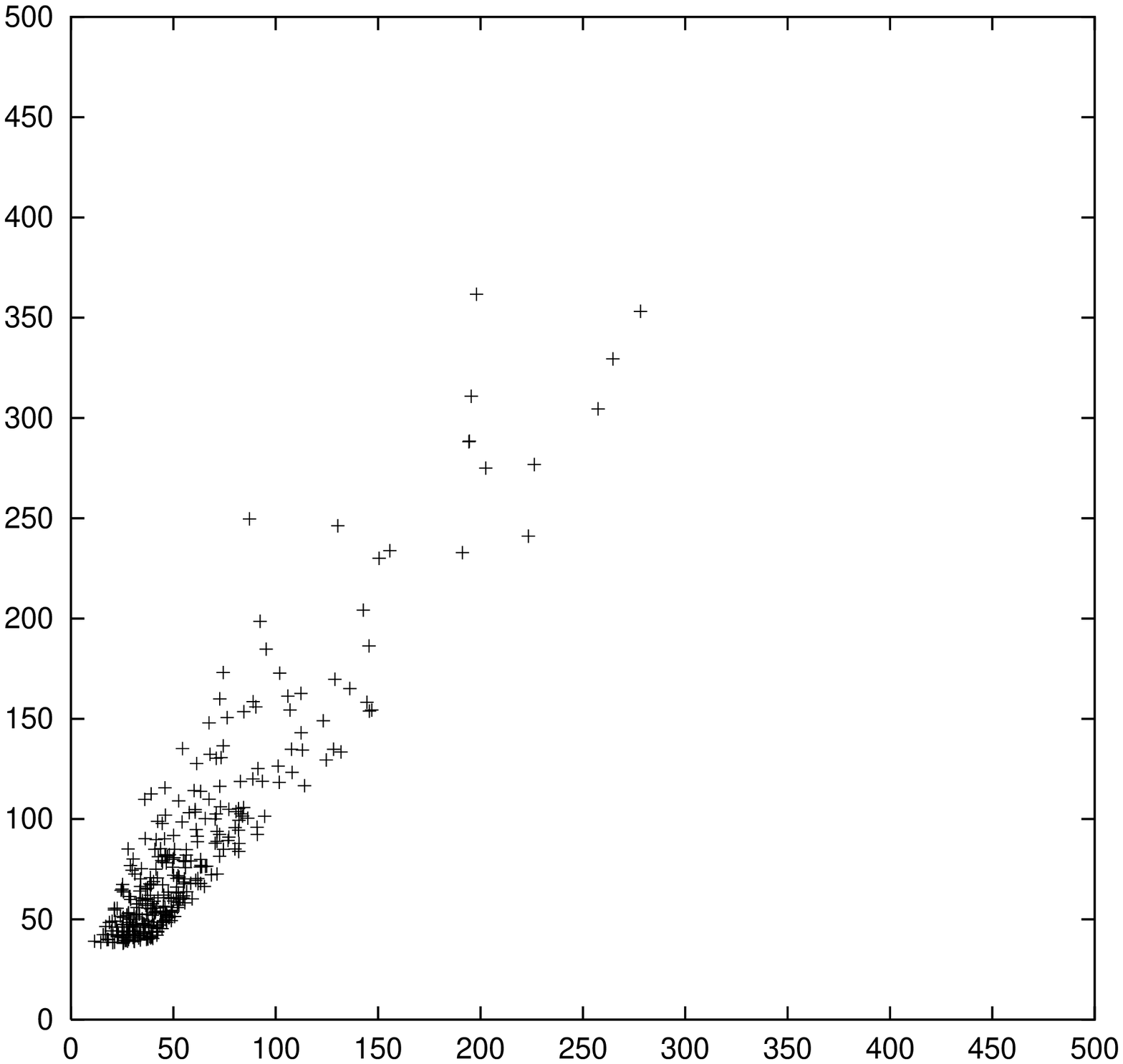}}
\end{picture}
\end{figure}

\end{document}